\pdfoutput=1
\documentclass[pra,twocolumn,amsmath,amssymb,superscriptaddress]{revtex4-2}
\usepackage{graphicx,amsmath,relsize,epstopdf,color,mathtools,bm,newtxtext,newtxmath,braket,rotating}
\usepackage[hyphenbreaks]{breakurl}
\usepackage[colorlinks=true,linkcolor=blue,citecolor=blue,urlcolor =blue]{hyperref}
\usepackage[normalem]{ulem}

\usepackage{booktabs}
\usepackage[table,xcdraw]{xcolor}
\newcommand{\eq}[1]{\begin{equation}\begin{aligned}#1\end{aligned}\end{equation}}

\newcommand{\Tr}{\mathop{\mathrm{Tr}} \nolimits}

\newcommand{\eu}{\mathrm{e}}
\newcommand{\iu}{\mathrm{i}}

\newcommand{\suggOLD}[1]{#1}
\newcommand{\sugg}[1]{{#1}}
\newcommand{\suggQ}[1]{#1}
\newcommand{\suggJPC}[1]{{#1}}
\newcommand{\suggJPCC}[1]{#1}
\usepackage{soul,xcolor}

\usepackage{dsfont}

\begin{document}
	\setstcolor{red}

	\title{Breaking the limits of purification: \sugg{Postselection} enhances heat-bath algorithmic cooling}
	
	\author{Aaron Z. Goldberg}
	\affiliation{National Research Council of Canada, 100 Sussex Drive, Ottawa, Ontario \suggQ{K1N 5A2}, Canada}
	\affiliation{Department of Physics, University of Ottawa, Advanced Research Complex, 25 Templeton Street, Ottawa,Ontario K1N 6N5, Canada}
	
	\author{Khabat Heshami}
	\affiliation{National Research Council of Canada, 100 Sussex Drive, Ottawa, Ontario \suggQ{K1N 5A2}, Canada}
	\affiliation{Department of Physics, University of Ottawa, Advanced Research Complex, 25 Templeton Street, Ottawa,Ontario K1N 6N5, Canada}
	\affiliation{Institute for Quantum Science and Technology, Department of Physics and Astronomy, University of Calgary, Alberta T2N 1N4, Canada}

	\begin{abstract}
		Quantum technologies require pure states, which are often generated by extreme refrigeration. Heat-bath algorithmic cooling is the theoretically optimal refrigeration technique: it shuttles entropy from a multiparticle system to a thermal bath, thereby generating a quantum state with a high degree of purity. Here, we show how to surpass this hitherto-optimal technique by taking advantage of \suggQ{a single binary-outcome measurement}. Our protocols can create arbitrary numbers of pure quantum states without any residual mixedness by using a recently discovered device known as a quantum switch \sugg{to put two operations in superposition}, \suggJPCC{with postselection certifying the complete purification}. 
	\end{abstract}
	
	\maketitle
	
	\section{Introduction} 
	Quantum technologies promise dramatic improvements in computation, communication, metrology, and more. The vast majority of protocols require access to pure quantum states, which are typically provided by extreme refrigeration of mixed quantum states, but refrigeration has intrinsic limits and cannot be applied in all physical scenarios. To wit, cooling of trapped-ion and nuclear-magnetic-resonance systems, among others, will always leave some residual mixedness that degrades the purity of the quantum states \cite{Schulmanetal1999,Boykinetal2002,Fernandezetal2004}.
	
\begin{figure*}
		\centering
		\includegraphics[width=0.75\textwidth]{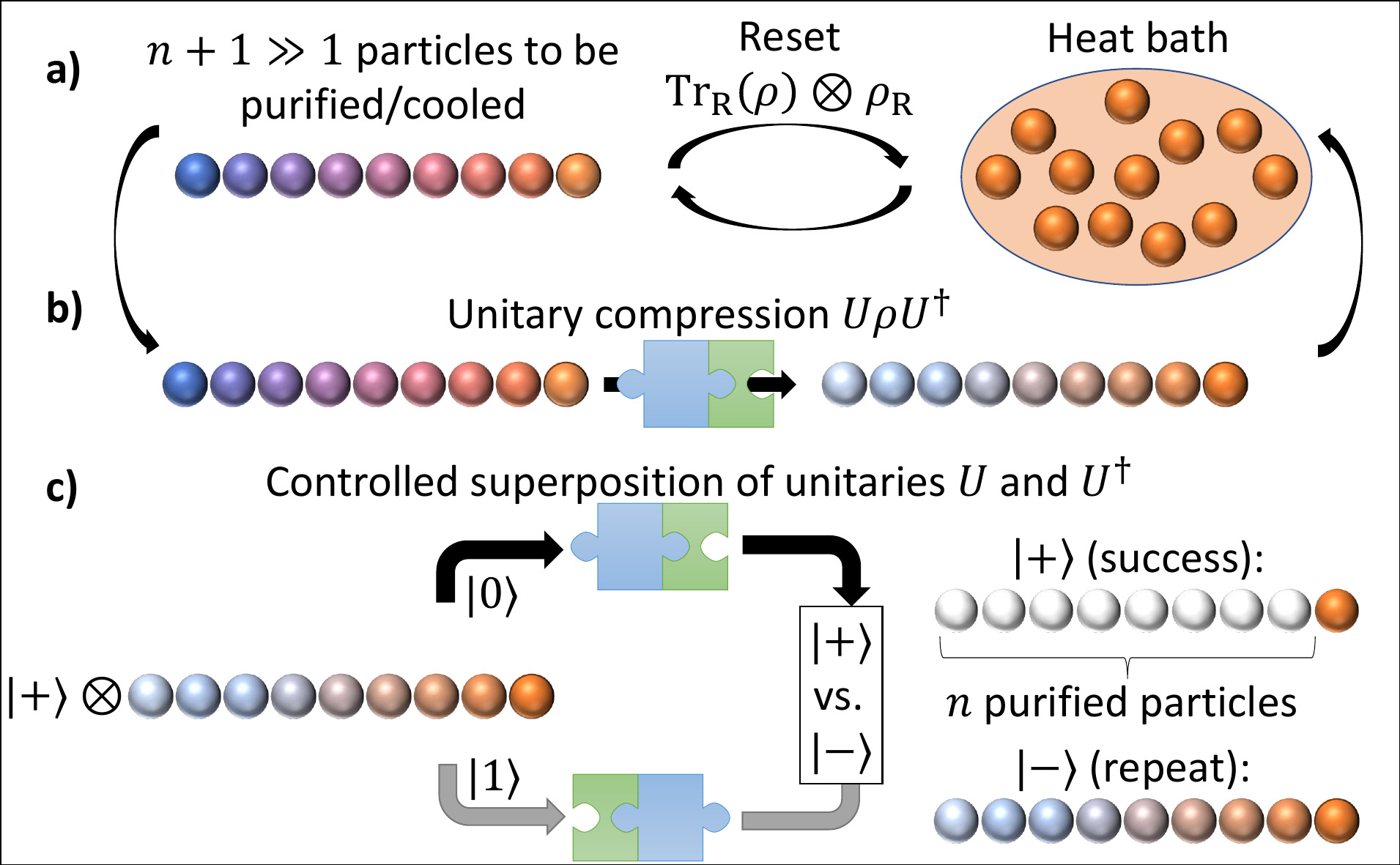}
		\caption{Schematic of HBAC augmented with postselection \suggQ{and coherent superpositions induced by the quantum switch}. Standard HBAC protocols repeatedly apply \textbf{a)} and \textbf{b)}; here, step \textbf{c)} \suggQ{is added} \suggOLD{after many cycles of steps \textbf{a)} and \textbf{b)}}. \textbf{a)} One of a series of particles is replaced by a particle that has thermalized with an external heat bath. \textbf{b)} A unitary operation shuttles entropy from the particles on the left to those on the right, thereby cooling the particles on the left. \textbf{c)} A control qubit in a superposition state causes two unitaries \sugg{$U$ and $U^\dagger$ to be applied to the particles in a superposition dependent on the state of the control}. Measuring the control to be $\ket{+}$ completely purifies all but one particle; measuring the control to be $\ket{-}$ requires a repetition of the protocol. The number of purified particles $n$ is unlimited and has negligible impact on the success probability. The probability of success can be increased arbitrarily by choosing appropriate unitaries $U$.
		}
		\label{fig:schematic}
	\end{figure*}

	Heat-bath algorithmic cooling (HBAC) was introduced as a method for simultaneously cooling a large ensemble of qubits, the most basic building blocks of quantum technologies. By serially replacing one of the qubits with one that was in thermal equilibrium with a large heat bath and performing a ``compression'' transformation on the qubits, as depicted in Fig. \ref{fig:schematic}\textbf{a-b)}, entropy can be shuttled to the bath to cool the multi-qubit system \cite{Schulmanetal1999,Boykinetal2002,Fernandezetal2004}. This idea has been comprehensively investigated \cite{Schulmanetal2005,RaeisiMosca2015,RodriguezBrionesetal2016,RodriguezBriones2017,Raeisietal2019,Alhambraetal2019,Pande2020arxiv,Raeisi2021,Farahmandetal2022} and experimentally demonstrated \cite{Baughetal2005,Zaiseretal2018arxiv}.
	
	There is a fundamental limit to the cooling and therefore the purity that HBAC can achieve \cite{Schulmanetal2005,RaeisiMosca2015,RodriguezBrionesetal2016}. This was recently shown to stem from assumptions about the unitarity of transformations on the multi-qubit systems \cite{Raeisi2021} and precludes any of the resulting states from ever being completely pure; instead, the optimal state is the probabilistic mixture given in Eq. \eqref{eq:old ultimate state} \suggOLD{\footnote{\suggOLD{If one assumes that the thermal bath is a resource whose dynamics can be manipulated, that one has access to \textit{a priori} knowledge of the states to be purified, and that the thermal bath is sufficiently cold, the extended HBAC methods presented in Ref. \cite{Alhambraetal2019} can increase the probability of creating a purified state.}}}. We here show that, by allowing for multiple unitary transformations to be applied in a superposition of their order, one can create completely pure states in a \textit{heralded} fashion. \sugg{A postselective measurement on a single control qubit allows us to unearth the absolutely pure component of a state with an unlimited number of qubits, reminiscent of the one clean qubit model of quantum computing \cite{KnillLaflamme1998}.}
	\suggOLD{These methods} simultaneously cool all of the qubits to their ground states, thereby \suggJPCC{certifiably circumventing} the limits that were previously imposed on HBAC. 
	
	\suggQ{That measurement may help with purification is trivial: projective von Neumann measurements should suffice for generating pure quantum states. However, this naive picture is too simplistic, as it requires $n$ ideal binary-outcome measurements to purify $n$ qubits. Alternative measurement-based purification schemes exist that require many measurements or only converge toward the purest state \cite{Eschneretal1995,Nakazatoetal2003,Nakazatoetal2004,Wuetal2004,Roaetal2006,Lietal2011,Machnesetal2012,Xuetal2014,Raoetal2016,Montenegroetal2018,Zhangetal2019,Pueblaetal2020,Yanetal2021}, such as a recent idea that proposes to perform $\gg n$ measurements on a single auxiliary qubit to purify $n$ qubits (with some finite success probability that decays exponentially with $n$) \cite{Konaretal2022} and one that seeks to reduce the number of measurements by using techniques from reinforcement learning \cite{YanJing2022}. In comparison, our method can \textit{completely} purify $n$ qubits with a \textit{single} binary measurement (and arbitrarily high success probability). This provides a significant advantage in any setting where measurements are challenging and realistic imperfections must be taken into account, which we address in this work.}
	
	\sugg{Coherently controlled superpositions} have become extremely relevant over the last decade. This phenomenon, \sugg{often studied through the lens of} indefinite causal order (ICO), has already been shown to break \textit{quantum} limits in computation \cite{Colnaghietal2012,Morimae2014,Araujoetal2014,Taddeietal2021}, communication \cite{Chiribella2012,Feixetal2015,Guerinetal2016,DelSantoDakic2018,Ebleretal2018,Procopioetal2019,Chiribellaetal2021}, metrology \cite{Mukhopadhyayetal2018arxiv,Frey2019,Zhaoetal2020,ChapeauBlondeau2021ICO}, and thermodynamics \cite{Guhaetal2020,Caoetal2022,ChenHasegawa2021arxiv,Guhaetal2022arxiv,Simonovetal2022}, each of which were already known to outperform \textit{classical} versions of the same protocols. ICO has been demonstrated in a number of groundbreaking experiments \cite{Procopioetal2015,Rubinoetal2017,Rubinoetal2022,Goswamietal2018,Weietal2019,Massaetal2019,Goswamietal2020,Guoetal2020,Rubinoetal2021}, many of which rely on a device known as a quantum switch to enable superpositions of the order in which unitary operations are applied to a system. Perhaps even more importantly, ICO provides a new approach to thinking about old problems and can inspire simpler solutions with definite causal order that take advantage of coherent superpositions \suggQ{of operations} to achieve the same goals \cite{ChapeauBlondeau2021ICOinspired,Nieetal2022arxiv}    \suggQ{\cite{ChiribellaZhao2022arxiv}}.
	This is a powerful new tool that is only beginning to be understood so, while it is remarkable to find yet another important application for \suggQ{coherent superpositions of unitaries}, it would not be surprising to discover that coherent-superposition protocols have many more varied applications that we are only beginning to fathom. 
	
	In our first protocol, we \suggOLD{apply HBAC until it converges to its optimal final state and then} replace the optimal unitary compression from HBAC with a controlled superposition of two unitaries that \sugg{each} achieve the original transformation, now depicted in Fig. \ref{fig:schematic}c). 
	\sugg{This replacement maintains the spirit of HBAC by adding a single auxiliary pure qubit to the compression step and not performing any measurements on the qubits to be purified, which is especially negligible in the limit of large numbers of qubits to be purified.}
	Measuring the control in a superposition basis can then herald the creation of an unlimited number of completely pure qubits. 
	We next explore and compare purification protocols that incorporate controlled operations with different unitaries to achieve even faster purification, which can be optimized in terms of the temperature of the heat bath and other physical resources. 
	\suggOLD{We find that an \sugg{\textit{unlimited}} number of pure states can be created with a \sugg{\textit{fixed}} number of input pure states with \sugg{\textit{any}} desired probability of success \sugg{arbitrarily close to $100\%$}, \sugg{easily justifying the introduction of auxiliary pure qubits and} making all of our techniques broadly applicable to quantum refrigeration and purification.}
	
    \suggQ{Coherent superpositions have already been studied in the context of cooling through ICO \cite{Caoetal2022,Nieetal2022arxiv,Nieetal2022inPress}, where quantum states can be serially cooled through subsequent applications of an ICO-cooling procedure. In contrast, our method achieves the ultimate cooling limit, that of pure states,  \suggJPC{by combining standard HBAC with} a single coherent superposition step. \suggJPCC{Standard HBAC achieves the majority of the cooling, then our method certifies completion of the cooling in a heralded manner.} This outperforms all other classical and quantum refrigerators and has implications for the role of coherence and measurement in thermodynamics.}

	\section{Heat-bath algorithmic cooling (HBAC)}
	The original techniques for HBAC have improved over the past few years. While the technique of Ref. \cite{Schulmanetal2005} is optimal in that it achieves the cooling limit \suggOLD{of Ref.} \cite{RaeisiMosca2015}, it is state dependent and thus complex to implement. It was superseded by the technique of Ref. \cite{Raeisietal2019}, which allows the ultimate cooling limit to be achieved without any knowledge of the state being cooled; we sketch this technique here [see Fig. \ref{fig:schematic} \textbf{a-b)}]. 
	
	The initial state to be cooled has $n+1$ qubits in a state represented by the density matrix $\rho$ \suggJPC{that need not take any particular form}. 
	Ultimately, the goal is to create the pure, ground state $\ket{\mathrm{\bf{g}}}=\ket{g}^{\otimes n+1}$,
	with the realistic goal being to create a state with as large of a probability as possible of being found in this state.
	We have access to a thermal bath at some temperature $T$ from which we can extract thermalized qubits with density matrices
	\begin{equation}
		\rho_{\mathrm{R}}=\frac{1}{z}\begin{pmatrix}
			\eu^{\varepsilon}&0\\0&\eu^{-\varepsilon}
		\end{pmatrix},
		\label{eq:reset qubit}
	\end{equation} while also being able to send any qubit to the bath such that it thermalizes into state $\rho_{\mathrm{R}}$\suggJPCC{, as is the standard assumption in HBAC \cite{Boykinetal2002,Fernandezetal2004,Schulmanetal2005,RaeisiMosca2015,RodriguezBrionesetal2016,RodriguezBriones2017,Raeisietal2019,Baughetal2005,Zaiseretal2018arxiv} and can be achieved through methods such as sympathetic cooling \cite{Barrettetal2003}}. Here, the subscript $\mathrm{R}$ stands for ``reset,'' the qubit is taken to be in the basis of ground and excited states $\ket{\mathrm{g}}$ and $\ket{\mathrm{e}}$ whose energies differ by $\suggJPCC{\Delta=}\sugg{2}\varepsilon k_B T>0$, $z=2\cosh\varepsilon$, and we have assumed the bath to be sufficiently large such that it quickly thermalizes on a timescale more rapid than any other relevant to the rest of the protocol \suggJPC{and thus can be used as an entropy repository without noticeably changing in temperature}. 
	
	First, whatever is occupying the position of the reset qubit is removed from the state and replaced with the state $\rho_{\mathrm{R}}$. This is mathematically represented by the operation [Fig. \ref{fig:schematic}\textbf{a)}]
	$\rho\to \Tr_{\mathrm{R}}\left(\rho\right)\otimes\rho_R$,
	where $\Tr_{\mathrm{R}}$ is the partial trace removing the qubit in the reset position. \suggJPC{This tends to send entropy to the bath because the reset qubit tends to have a higher temperature and entropy than the bath qubits.} Extensions to HBAC that allow access to the internal dynamics of the bath \cite{Alhambraetal2019} will be addressed later \suggQ{and essentially allow one to completely cool the entire system by allowing arbitrary quantum channels to be imposed on a state, which is an unlimited resource for state preparation \cite{Wuetal2007}}.
	Next, a unitary operation [Fig. \ref{fig:schematic}\textbf{b)}; we are using the computational basis]
	\begin{equation}
		U=\mathrm{Diag}(
		1,\underbrace{\sigma_x,\cdots,\sigma_x}_{2^n-1\,\mathrm{times}}
		,1
		),
	\end{equation}
where $\sigma_x$ is a Pauli matrix and all of the off-diagonal entries are null,
	acts on the state to shift the larger probabilities like $\eu^{\varepsilon}$ toward the $\ket{\mathrm{\bf{g}}}$ part of the system and the smaller probabilities like $\eu^{-\varepsilon}$ toward the opposite end\suggJPC{, thereby concentrating the entropy toward the reset qubit}. This "two-sort" unitary can be efficiently implemented \cite{Raeisietal2019} \suggOLD{using techniques from Ref. \cite{SaeediPedram2013} or the more recent techniques of Ref. \cite{Bakeretal2021arxiv}}. The entire process is then repeated until, with a high degree of probability, the state reaches the theoretically coldest (most pure) state possible \suggJPC{with HBAC; as much entropy as possible with this method has been pumped from the state into the thermal bath}. \suggQ{The larger $n$ is, the more challenging $U$ is to implement and the more steps are required until the protocol converges.}
	
	Only the diagonal elements of the density matrix are relevant to evaluating the success of this protocol. Denoting these by the $2^{n+1}$-component vector $\boldsymbol{\lambda}^t$ after the $t$th iteration of the protocol, $\lambda_1^t$ represents the probability of finding the state to be $\ket{\mathrm{\bf{g}}}$, $\lambda_2^t$ the probability of all of the qubits being in their ground states except for the reset qubit, and so on. Alternatively, we can focus on the probability distribution for the $n$ qubits after ignoring the reset qubit, now denoted by the $2^n$-component vector $\boldsymbol{p}^t$. In terms of these components, the removal of the reset qubit ensures
	$p^t_k=\lambda^t_{2k-1}+\lambda^t_{2k}$ ($1\leq k\leq 2^n$), the addition of a thermal state leads to the intermediary probabilities
	$\lambda^{t^\prime}_{2k-1}=p_k^t \eu^{\varepsilon}/z$ and $\lambda^{t^\prime}_{2k}=p_k^t \eu^{-\varepsilon}/z$ ($1\leq k\leq 2^n$), and the two-sort unitary rearranges the resulting vector to obey $\lambda^{t+1}_{2k}=\lambda^{t^\prime}_{2k+1}$ and $\lambda^{t+1}_{2k+1}=\lambda^{t^\prime}_{2k}$ ($1\leq k<2^n$). Overall, the probabilities are updated according to the rules $\boldsymbol{p}^{t+1}=\boldsymbol{T}\boldsymbol{p}^t$, with 
	transfer matrix
	\begin{equation}
		\boldsymbol{T}=\frac{1}{z}\begin{pmatrix}
			\eu^{\varepsilon} & \eu^{\varepsilon}&0&\cdots&0\\
			\eu^{-\varepsilon} & 0&\eu^{\varepsilon}&\cdots&0\\
			0 & \eu^{-\varepsilon}&0&\cdots&0\\
			0 & 0&\cdots&\ddots&\vdots\\
			0 & 0&\cdots&\eu^{-\varepsilon}&\eu^{-\varepsilon}\\
		\end{pmatrix}.
	\end{equation}

	Because $\boldsymbol{T}$ is almost a tridiagonal Toeplitz matrix, other than the first and last entries on its main diagonal, its eigenvalues can readily be computed. It has a single dominant eigenvalue corresponding to the hitherto-ultimate eigenstate
	\begin{equation}
		\boldsymbol{p}^\infty=\frac{1-\eu^{-2\varepsilon}}{1-\eu^{-(2^n)2\varepsilon }} \left(1,\eu^{-2\varepsilon},\eu^{-4\varepsilon},\cdots,\eu^{-(2^{n} -1)2\varepsilon}\right)^\top,
		\label{eq:old ultimate state}
	\end{equation} which is denoted by the superscript $\infty$ because it is the ultimate final state toward which HBAC protocols evolve \suggJPC{regardless of the initial state $\rho$}.
	Notably, none of the qubits is completely pure: The purity of the $j$th qubit $\mathrm{Tr}(\rho_j^2)$ ranges from $\frac{\eu^{\varepsilon 2^{n+1}}+1}{\left(\eu^{\varepsilon 2^n}+1\right)^2}\approx 1-2\eu^{-2^{n+1}\varepsilon}$ for the most pure qubit to $\frac{1}{2} \left(1+\tanh ^2\varepsilon\right)\approx \frac{1+\epsilon^2}{2}$ for the least pure qubit\suggJPC{, while a pure qubit has $\mathrm{Tr}(\rho_j^2)=1$}. \suggJPC{These can also be expressed in terms of effective temperatures for the $j$th qubit through the standard relation \suggJPCC{$T_j=2\varepsilon T/\cosh^{-1}\tfrac{\mathrm{Tr}(\rho_j^2)}{1-\mathrm{Tr}(\rho_j^2)}$}, which show the most pure qubit to be at the tiny temperature $T/2^{n-1}$ and the least pure qubit to be at the temperature of the thermal bath. Pure qubits have effective temperature $T_j=0$ and maximally mixed qubits have effective temperature $T_j\to\infty$. Performing HBAC for an \textit{infinite} amount of time will never supersede these purity and temperature limits.}
	This is the benchmark that HBAC protocols employing controlled superpositions are presently shown to beat.

	\section{The Quantum Switch}
	\color{black}
	ICO leads to intriguing questions about the nature of causality \cite{Hardy2007,Oreshkovetal2012,Araujoetal2015,BaumelerWolf2016,OreshkovGiarmatzi2016,Zychetal2019,Oreshkov2019,Dimicetal2020,Milzetal2021}. One of its offshoots is the development of a ``quantum switch,'' whose theoretical motivation \cite{Colnaghietal2012,Chiribellaetal2013,Friisetal2014,Friisetal2015,Ramboetal2016,Giacominietal2016} and subsequent experimental implementation \cite{Procopioetal2015,Rubinoetal2017,Rubinoetal2022,Goswamietal2018,Weietal2019,Massaetal2019,Goswamietal2020,Guoetal2020,Rubinoetal2021} have pushed the capabilities of quantum technologies well beyond what can be achieved without it \cite{GoswamiRomero2020}. We will show that a quantum switch is the only tool that we need in order to outperform HBAC protocols. In fact, the essential feature of quantum switches is that they create controlled superpositions of unitary operations, which leads us to discover other protocols with definit\suggQ{e} causal order that also herald perfect purification.
	
	A quantum switch allows a single ``control'' state to control the order in which quantum operations are applied to a target state. For example, a two-switch can be represented by the unitary operator
	\begin{equation}
		U_{\mathrm{ICO}}=\ket{0}_{\mathrm{control}}\bra{0}\otimes U_B U_A +\ket{1}_{\mathrm{control}}\bra{1}\otimes U_A U_B:
	\end{equation} $U_A$ is applied before $U_B$ when the control state is $\ket{0}$ and $U_B$ is applied before $U_A$ when the control state is $\ket{1}$ \suggOLD{\cite{Chiribellaetal2013}}.
	We define the superposition states
	$
	\ket{\pm}=\frac{\ket{0}\pm\ket{1}}{\sqrt{2}}.
	$  If the control system is prepared in $\ket{+}$ \suggOLD{and} the unitary is applied, \suggOLD{the state of the control will be entangled with the state of the system.} \suggOLD{Then, if} the control is measured to be $\ket{\pm}$, \suggOLD{which occurs with probability $P_\pm$, }the target system will evolve as
	\begin{equation}
	\begin{aligned}
		\rho\to& \left(U_A U_B\rho U_B^\dagger U_A^\dagger
		+U_B U_A\rho U_A^\dagger U_B^\dagger\right.\\
		&\left.
		\pm U_A U_B\rho U_A^\dagger U_B^\dagger\pm U_B U_A\rho U_B^\dagger U_A^\dagger\right)/4\suggOLD{P_\pm}.
		\label{eq:quantum switch}
		\end{aligned}
	\end{equation} The first two terms are the convex combinations of the two orders of unitaries being applied, while the final two terms are unique interference effects stemming from ICO. These interference effects are what allow ICO to outperform standard quantum methods for a growing number of protocols. Since they stem from interference and postselection, we show how to achieve our results using definite causal order and merely taking advantage of interference and postselection.

	\section{The Quantum Switch and Postselection dramatically improve HBAC}
	\color{black}

	Putting together the pieces, we consider HBAC augmented with \suggQ{controlled superpositions and postselection} (``HBAC+\suggQ{PS}'' in Table \ref{tab:my-table}): after applying standard HBAC and exhausting its usefulness \suggJPC{to create the state given in Eq. \eqref{eq:old ultimate state}}, we replace the unitary transformation $U$ with a pair of unitary transformations $U_A$ and $U_B$ whose order of application is controlled by a single auxiliary qubit (Fig. \ref{fig:schematic}). We choose $U_A$ and $U_B$ such that both of the combined unitaries $U_A U_B$ and $U_B U_A$ independently lead to the transformation matrix $\boldsymbol{T}$ \suggOLD{to avoid straying from the assumptions of HBAC}, with the interference effects arising only from ICO through Eq. \eqref{eq:quantum switch}. These are defined by
	\begin{equation}
	\begin{aligned}
		%
		U_A&=\mathrm{Diag}(
		1,
		\sigma_y,\cdots,\sigma_y,1
		),\\
		U_B&=\mathrm{Diag}(
		1,
		\underbrace{\sigma_z,\cdots,\sigma_z}_{2^n-1\,\mathrm{times}},1
		).
		\label{eq:U1 and U2 original}
		\end{aligned}
	\end{equation}
	\sugg{By choosing unitaries $U_A$ and $U_B$ that are fixed \textit{a priori}, we do not require a generic quantum switch that will allow for a controlled superposition of arbitrary pairs of unitaries, such that our protocols can equally be achieved using \textit{any} controlled operation that yields $U_{\mathrm{ICO}}$; implementing $U=U_BU_A$ and $U^\dagger=U_A U_B$ conditional on the state of the control being $\ket{0}$ and $\ket{1}$, respectively, will suffice.} \suggQ{This allows us to take techniques \textit{inspired by} ICO to devise new protocols with fixed causal structures that significantly outperform previous protocols.}
	
	Because $\sigma_y\sigma_z=\iu \sigma_x$ and $\sigma_z\sigma_y=-\iu \sigma_x$, many cancellations result from Eq. \eqref{eq:quantum switch}. The probabilities now rearrange as
	$\lambda^{t+1}_{1}=\lambda^{t^\prime}_{1}(1\pm 1)/2$,
	$\lambda^{t+1}_{2^{n+1}}=\lambda^{t^\prime}_{2^{n+1}}(1\pm 1)/2$, and
	$\lambda^{t+1}_{2k}=\lambda^{t^\prime}_{2k+1}(1\mp 1)/2$ and $\lambda^{t+1}_{2k+1}=\lambda^{t^\prime}_{2k}(1\mp 1)/2$ for $1\leq k< 2^n$. 
	When this quantum switch is added to the above HBAC protocol and the control is measured to be $\ket{+}$, the transfer matrix becomes \suggJPC{(see Appendix \ref{app:transfer matrix} for a didactic calculation)}
	\begin{equation}
		\boldsymbol{T}_+\suggJPC{\propto}\frac{1}{z}\mathrm{Diag}(
		\eu^{\varepsilon},\underbrace{0,\cdots,0}_{2^n-2\,\mathrm{times}},e^{-\varepsilon}
		).
		\label{eq:T+}
	\end{equation}
	Simply measuring one of the \suggOLD{target system's $n$} qubits in the ground/excited-state basis collapses the remaining $n-1$ qubits to either \suggJPC{the completely pure} state $\ket{\mathrm{\bf{g}}}$ or the similarly-defined \suggJPC{completely pure} state $\ket{\mathrm{\bf{e}}}$ \suggOLD{(this step is not depicted in Fig. \ref{fig:schematic} because we will see that it can be circumvented)}. Since the result is directly known, the resulting state has been completely purified, and a simple ``$\pi$ pulse'' can be applied to each qubit to deterministically enact the transformation $\ket{\mathrm{\bf{e}}}\to\ket{\mathrm{\bf{g}}}$. In fact, because only $\lambda_1$ and $\lambda_{2^{n+1}}$ are nonzero when the control is \suggOLD{measured to be} $\ket{+}$, this creates $n$ pure qubits\suggJPC{, each with purity $\Tr(\rho_j^2)=1$ and effective temperature $T_j=0$}. This thus achieves the ultimate goal of HBAC without being limited by the final state of Eq. \eqref{eq:old ultimate state}.
	
	The probability of measuring the control to be $\ket{+}$ tends quickly toward $P\suggOLD{_+}=\left(1-\eu^{-2\varepsilon}\right)\eu^{\varepsilon}/z\approx \varepsilon$ for moderate-to-large $n$\suggQ{, similar to the probability of success in Ref. \cite{Pyshkinetal2016}}. \suggJPCC{This probability is independent of $n$ such that any number of pure qubits can be created without decreasing the probability of success and without increasing the number of qubits that need to be measured.} The extra thermodynamic cost for increased $n$ only comes via the standard increased HBAC cost of achieving the state given in Eq. \eqref{eq:old ultimate state}; regardless of $n$, a single positive measurement will suffice to purify an unlimited number of states \suggJPC{to each have purity $1$ and effective temperature $0$}. For completeness, we provide the transfer matrix for the case when the control is measured to be $\ket{-}$:
	$\boldsymbol{T}_-=\boldsymbol{T} - \boldsymbol{T}_+$. If this result is obtained, the overall process can simply be repeated until the control qubit is measured to be $\ket{+}$, such that the success of the protocol is unequivocally known.
	
	How can our protocol overcome what was previously thought to be a fundamental limit? 
	 \cite{Wuetal2013,TicozziViola2014,Silvaetal2016,Raeisi2021}.
	 \suggJPCC{Simply supplying a pure auxiliary qubit to a standard HBAC protocol does not achieve the same goal, because that can only be used to create a single pure qubit \cite{Wuetal2013,TicozziViola2014,Silvaetal2016,Raeisi2021}.}
	The key is that, even though $U_{\mathrm{ICO}}$ is unitary, our protocol involves a \suggOLD{single} measurement step that allows us to avoid the assumptions of Ref. \cite{Raeisi2021}. Adding a measurement step to standard HBAC would not help: since the final state is the probabilistic mixture given in Eq. \eqref{eq:old ultimate state}, no measurement result will \sugg{\textit{herald}} the rest of the qubits being pure. It is the combination of \sugg{coherent superpositions of unitaries controlled by a single auxiliary qubit} with the measurement \sugg{of that single qubit}\suggOLD{, which we argue is a minimal addition to HBAC that provides significant benefits,} that allows us to herald the creation of \textit{unlimited} pure-qubit states.
	
	\suggQ{One could argue that performing a projective measurement on each of the $n$ qubits from Eq. \eqref{eq:old ultimate state} would suffice to purify the entire system. Such a procedure, however, comes at a steep cost, including caveats such as: a) If the measurements are in any way destructive, the purified qubits will be degraded or worthless; \suggJPC{in contrast, our measurements are performed on auxiliary qubits, which need not be preserved;} b) The apparatus required to perform a large number of measurements is certainly more complicated than that required to perform a single measurement, with the former perhaps requiring synchronization to measure and purify all $n$ qubits simultaneously or quantum memories to store the purified qubits; and c) An imperfect measurement that leaves the measured qubit in the desired state with probability $1-\delta$ will provide us with $n$ qubits in their ground states with probability $1-\delta$, in contrast to the serial measurement protocol that could only achieve the same result with an exponentially poorer probability $(1-\delta)^n$. Together, these show that only \textit{judicious} measurements can be used for state purification.}
	
	\suggJPC{When the postselection fails too many times, one has to repeat the HBAC procedure before trying the controlled superposition again, which is costly. Fortunately, we can find related protocols that increase the probability of postselection success by an arbitrary amount, such that one can perform HBAC once, exhaust its usefulness, then use the quantum switch once to herald the complete purification of the vast majority of the states with arbitrarily high success probability. We turn to these ideas next.}

	\section{Refining the idea}
	The key to the advantage of \suggQ{postselection} comes from the cancellations conferred by $U_{\mathrm{ICO}}$ relative to the regular sorting unitary $U$. We have introduced these cancellations as replacing a single step in the HBAC protocol, but they can, in fact, replace the entire protocol. As well, different unitaries $U$ can dramatically increase the probability of successful postselection. We shed light on some of the many avenues down which one can proceed after adding quantum switches to their toolbox.

	\begin{table}[]
		\centering
		\caption{Resources required and probability of success of purification schemes using \sugg{postselection}. \suggJPCC{The first resource is a thermal bath with a fixed temperature $T$ that can dissipate entropy from any qubit put in contact with it until the qubit reaches temperature $T$. For a qubit with energy gap $\Delta$, this resource is quantified by $\varepsilon=\Delta/2K_BT$.} Colder bath temperatures lead to larger values of $\varepsilon$. \suggJPCC{The second resource is the number of auxiliary pure qubits required to power the cooling algorithm. Most of our algorithms only require one such auxiliary pure qubit, while the PS tree sort method can have $n^*$ reduced to 1 if quantum nondemolition measurements are used. The number of output pure qubits informs us of the usefulness of the algorithm, telling us how many qubits the algorithm will certify to be pure. The probability of success $P_+$ dictates how many times the algorithm will have to be repeated, on average $\mathcal{O}(1/P_+)$ times, in order to obtain that number of output pure qubits. A full repetition of the algorithm for HBAC, HBAC+PS, and HBAC+$k$PS can require full or partial HBAC to recreate an initial state close to that of Eq. \eqref{eq:old ultimate state}.} \suggJPC{The phrase ``pure qubit'' refers to having purity $\Tr(\rho_j^2)=1$ and is equivalent to \suggJPCC{a qubit having} effective temperature $T_j=0$. \suggJPCC{ There is no upper limit to the integer $n$; HBAC+$k$PS is most useful because the probability of success can be increased exponentially with $k$ while creating an unlimited number $n+1-k$ pure qubits}.}}
		\label{tab:my-table}
		\begin{tabular}{@{}ccccc@{}}
			\toprule
			Scheme        & Bath          & Input pure qubits & Output pure qubits & $P\suggOLD{_+}$                                 \\ \midrule
			HBAC          & $\varepsilon$ & 0                 & 0                  & 1                                   \\ 
			HBAC+\suggQ{PS}      & $\varepsilon$ & 1                 & $n$              & $\varepsilon$                      \\
			\suggQ{PS} tree sort & none          & $n^*$           & $n$              & 1                                   \\
			\suggQ{PS} alone     & none          & 1                 & $n$                & \suggOLD{$\lambda^0_{1}+\lambda^0_{2}$} \\
			HBAC+$k$\suggQ{PS}   & $\varepsilon$ & 1                 & $n+1-k$              & $2^{k}\varepsilon$                \\ \bottomrule
		\end{tabular}
	\end{table}
	
	\suggOLD{First}, consider generalizing the superposed unitaries $U_A$ and $U_B$ by changing the locations of the Pauli matrices on their diagonals. If we choose 
	\begin{equation}
		U_A=\mathrm{Diag}(\underbrace{1,\cdots,1}_{2^n \,\mathrm{times}},\underbrace{\sigma_y,\cdots,\sigma_y}_{2^{n-1} \,\mathrm{times}})
	\end{equation} and similarly for $U_B$ with $\sigma_z$ replacing each $\sigma_y$, we find the transfer matrices \begin{equation}
	\begin{aligned}
		\boldsymbol{S}_+&\suggJPC{\propto}\mathrm{Diag}(\underbrace{1,\cdots,1}_{2^{n} \,\mathrm{times}},\underbrace{0,\cdots,0}_{2^{n} \,\mathrm{times}}),\\
		\boldsymbol{S}_-&\suggJPC{\propto}\mathrm{Diag}(\underbrace{0,\cdots,0}_{2^n \,\mathrm{times}},\underbrace{\sigma_x,\cdots,\sigma_x}_{2^{n-1} \,\mathrm{times}})
		\end{aligned}
	\end{equation} 
	\suggOLD{for the evolution of the coefficients of the entire state $\boldsymbol{\lambda}^{t+1}=\boldsymbol{S}_{\pm}\boldsymbol{\lambda}^t$ conditioned on measuring the control to be in state $\ket{\pm}$.}
	Since these transformations hold for arbitrary permutations of the locations of the Pauli matrices on the diagonals of $U_A$ and $U_B$, this process can be repeated $n$ times so that the resulting density matrix will be guaranteed to have $n$ pure qubits and one mixed qubit, which we call the ``\suggQ{PS} tree sort'' method (Table \ref{tab:my-table}). However, this latter process requires $n$ pure qubits to enable the $n$ applications of the quantum switch, so it may not be physically useful \footnote{\sugg{This latter process is only useful if the control qubit can be measured nondestructively such that a single pure qubit can be recycled for all $n$ processes.}}\suggJPCC{.} 
	
	We thus see that applying \sugg{controlled unitaries} to upgrade HBAC protocols requires nuance. At least one pure qubit must be used to enable \suggQ{postselection}, while the goal of HBAC is to generate a large number of pure qubits. It then follows that one can tailor an HBAC+\suggQ{PS} setup to optimize the number of pure qubits generated for a given number of input pure qubits, desired probability of success, and temperature of the heat bath, among other parameters.
	
	For a given initial $n+1$-qubit state, the ideal single-quantum-switch process uses $U_A$ and $U_B$ similar to that of Eq. \eqref{eq:U1 and U2 original}:
	\begin{equation}
	\begin{aligned}
		U_A&=\mathrm{Diag}(1,1,\sigma_y,\cdots,\sigma_y),\\
		U_B&=\mathrm{Diag}(1,1,\underbrace{\sigma_z,\cdots,\sigma_z}_{2^{n}-1 \,\mathrm{times}}).
		\label{eq:U1 and U2 ideal}
		\end{aligned}
	\end{equation}
	Then, the \sugg{coherent superposition} step alone \suggOLD{(``\suggQ{PS} alone'' in Table \ref{tab:my-table})} yields $n$ pure qubits after measuring or simply ignoring the \suggOLD{reset} qubit, conditional on finding the control qubit to be $\ket{+}$. \suggOLD{This is depicted in Fig. \ref{fig:schematic}\textbf{c)}, where the reset qubit may be ignored at the end of the entire protocol.} What is the probability of success for this protocol? It is $\lambda_1^0+\lambda_2^0$, the same as the probability of the original system to have all of the final $n$ qubits be in their ground states. For a large number of qubits, this probability may be low, but the measurement can be repeated indefinitely until it is successful.
	
	HBAC can help here (``HBAC+1\suggQ{PS}'' in Table \ref{tab:my-table}): after performing HBAC with a large number of qubits, the probability distribution tends toward $\boldsymbol{p}^\infty$, with which the probability of \sugg{purification} success becomes double that of ``HBAC+\suggQ{PS}:''
	\begin{equation}
		P\suggOLD{_+}=\lambda_1^\infty+\lambda_2^\infty=p_1^\infty\approx 2\varepsilon
		,\quad \frac{1}{2^{n+1}}
		\ll \varepsilon\ll 1.
	\end{equation} One need only repeat the process $m=\mathcal{O}(1/ \varepsilon)$ times in order to achieve any desired success probability $P_{\mathrm{des}}$, which can be used to make an \textit{unlimited} number $n\gg m$ pure qubits. Small $\varepsilon$ corresponds to a hot heat bath and closely spaced energy levels $\ket{\mathrm{g}}$ and $\ket{\mathrm{e}}$, so this allows for significant refrigeration even with a hot bath\suggJPCC{.} This works because the probability of success remains approximately constant with increasing $n$. \sugg{For example, $\varepsilon\sim 10^{-5}$ at room temperature \cite{Schulmanetal1999} requires a supply of $\mathcal{O}(10^5)$ pure qubits to create, for example, $\mathcal{O}(10^{10^{10\cdots}})$ pure qubits.} If, instead, one has access to a cold thermal bath \sugg{around $0.1 \,\mathrm{K}$}, with large $\varepsilon\sugg{\sim\mathcal{O}(0.1)}$ \sugg{\cite{Schulmanetal1999}}, the probability of success tends to unity with \sugg{$\mathcal{O}(10)$} protocol\sugg{s}, so only \sugg{ten} pure qubit\sugg{s} need be supplied to generate an immense number $n\sugg{\sim\mathcal{O}(10^{10^{10\cdots}})}$ pure qubits.
	
	In a regime with miniscule $\varepsilon$, one can employ other tricks to improve the protocol. One can create a quantum switch with the unitaries from Eq. \eqref{eq:U1 and U2 ideal} replaced by unitaries with $2^{k}$ ones placed on the main diagonal before the remaining $2^{n}-2^{k-1}$ Pauli matrices for some integer $k$. Successfully measuring the control to be $\ket{+}$ and ignoring the final $k$ qubits will then yield $n+1-k$ pure qubits in their ground states (``HBAC+$k$\suggQ{PS}'' in Table \ref{tab:my-table}). Combined with the HBAC protocol, this will succeed with probability
	\begin{equation}
		P\suggOLD{_+}=\frac{1-\eu^{-\varepsilon 2^{k}}}{1-\eu^{-\varepsilon 2^{n+1}}}
		\approx 2^{k}\varepsilon\quad 2^{k}\varepsilon\ll 1.
		\label{eq:arbitrarily large prob}
	\end{equation} Even with $k\ll n$, the leading-order term $2^k\varepsilon$ can significantly hasten the convergence to a successful result. This now only requires $m=\mathcal{O}(2^{-k}\varepsilon^{-1})$ trials to succeed in creating $n+1-k\gg m$ pure qubits. \sugg{For example, with room temperature $\varepsilon\approx 10^{-5}$, purifying $k=20$ fewer qubits out of any large number $n$ will have a success probability greater than $99.99\%$ for a single input pure qubit}\suggJPCC{.} \suggQ{This is in stark contrast to any other method that cools a system through a single measurement, where the probability cannot be increased past $\epsilon$ \cite{Pyshkinetal2016}.} Many such tricks can further expedite HBAC using \suggQ{controlled superpositions and postselection}, thus establishing the import of quantum switches and related technologies to cooling protocols.
	
	\section{Controlled superpositions inspired by ICO}	
	\suggOLD{Finally, \suggQ{we show even more explicitly how} ICO and the quantum switch inspire other heralded methods using \textit{definite} causal order to break the limits of HBAC. \suggQ{When we designed our protocols above, we started by breaking the optimal unitary from standard HBAC into two parts such that applying either $U=U_B U_A$ or $U^\dagger=U_A U_B$ would achieve the same result as standard HBAC. Now, instead, we design completely new unitaries that have the same interference properties as $U$ and $U^\dagger$ but bear no resemblance to the original unitary from HBAC.} The same results \suggQ{as our above protocols} can be obtained, for example, by using controlled superpositions of the form $V=\ket{0}\bra{0}\otimes v+\ket{1}\bra{1}\otimes v^\dagger$, where $v=\mathrm{Diag}(1,1,\iu,\cdots,\iu)$, with the interference again coming from postselection. We can use similar tricks to increase the probabilities of success to match that of HBAC+$k$\suggQ{PS} by changing the number of $1$s before the remaining $\iu$s on the diagonal of $v$. 
	} \sugg{These postselective controlled superposition methods inspired by ICO can be much simpler to implement in practice and can readily be used to outperform HBAC.}

In this section, we provide an explicit scheme for an augmented HBAC protocol that is inspired by the quantum switch, which, to the best of our knowledge, has never before been demonstrated.
The unitaries we would like to implement take the form
\eq{
		V_k=\ket{0}_{\mathrm{control}}\bra{0}\otimes v_k +\ket{1}_{\mathrm{control}}\bra{1}\otimes v_k^\dagger,
		\label{eq:control superposition}
	}where 
	\eq{v_k=\mathrm{Diag}(\underbrace{1,\cdots,1}_{2^k\,\mathrm{times}},\underbrace{\iu,\cdots,\iu}_{2^{n+1}-2^k\,\mathrm{times}}).\label{eq:vk unitary}
	} Because we are not restricted to $2^{n+1}-2^k$ being an even number, $k$ can be as small as $0$, unlike the protocols in the main text that are restricted to $k\geq 1$.
	
\begin{figure*}
		\includegraphics[width=0.75\textwidth,trim={30 80 0 80},clip]{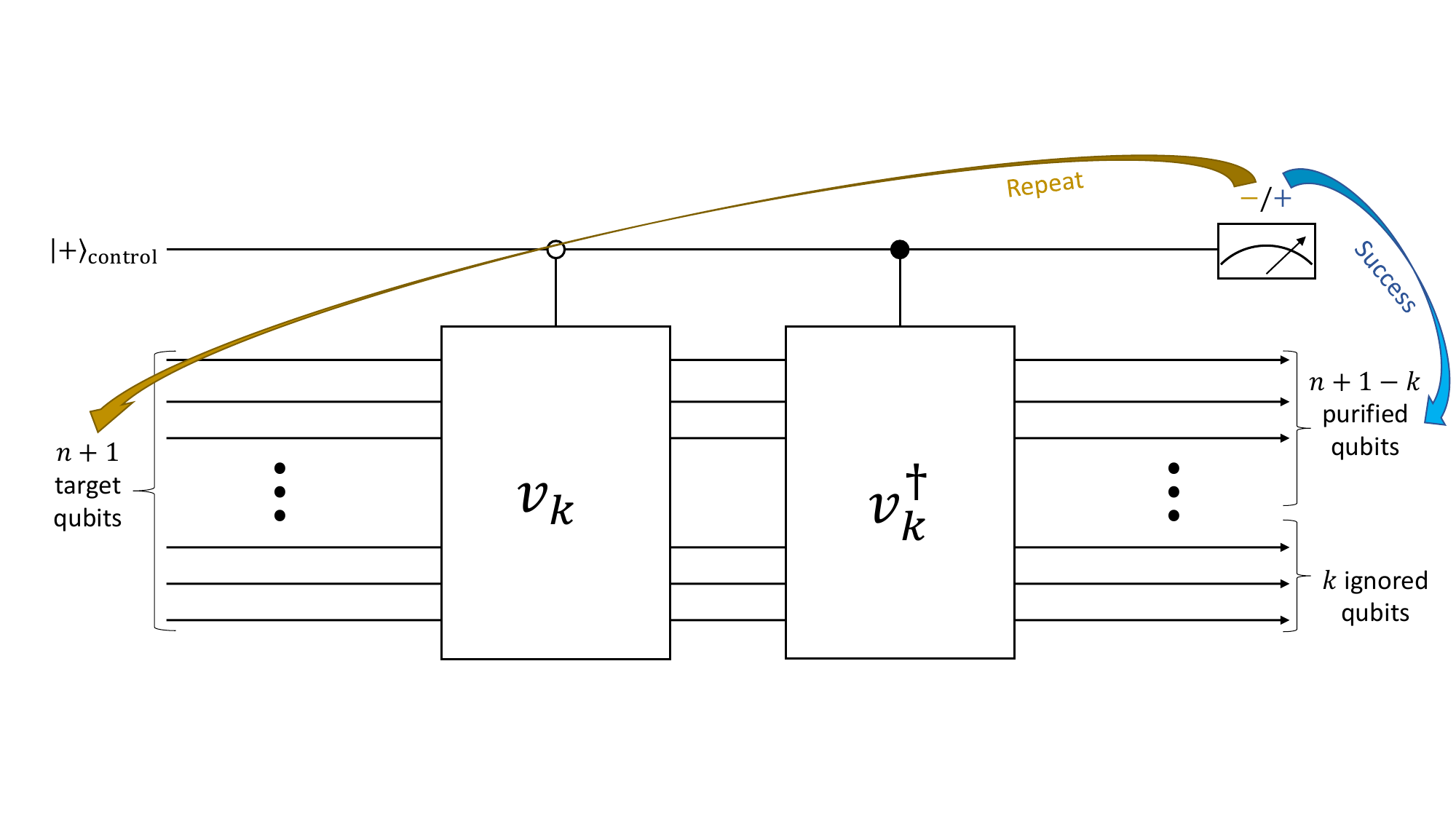}\caption{
			Quantum circuit diagram for the creation of $n+1-k$ pure qubits using a single pure control qubit. The unitary $v_k$ ($v_k^\dagger$) is applied to the target qubits when the control is in state $\ket{0}$ ($\ket{1}$) such that a superposed control state $\ket{\pm}=(\ket{0}\pm\ket{1})/\sqrt{2}$ yields an overall entangled state. When the control is measured to be in $\ket{+}$ at the end, the first $n+1-k$ qubits are guaranteed to be pure. The specific unitaries $v_k$ are depicted in Fig. \ref{fig:circuit diagram V}.}\label{fig:circuit diagram control}
	\end{figure*}

	The first step to create the unitaries is a regular controlled operation, as depicted in Fig. \ref{fig:circuit diagram control}. There, the state of the control qubit dictates which of the two unitaries $v_k$ or $v_k^\dagger$ is applied to the target system, enacting Eq. \eqref{eq:control superposition}. A control in the superposition state then yields entanglement between the control state and the unitarily evolved target state. As in the HBAC+$k$\suggQ{PS} protocol, successfully measuring the control to be in state $\ket{+}$ and ignoring the final $k$ qubits will herald the creation of $n+1-k$ completely purified qubits. This happens with the arbitrarily large probability given in Eq. \eqref{eq:arbitrarily large prob}. Measuring the control to be in state $\ket{-}$, which happens with probability $1-P_+$ dwindling exponentially with $k$, requires a repetition of the protocol.

\begin{figure*}
		\includegraphics[width=0.75\textwidth,trim={80 140 80 140},clip]{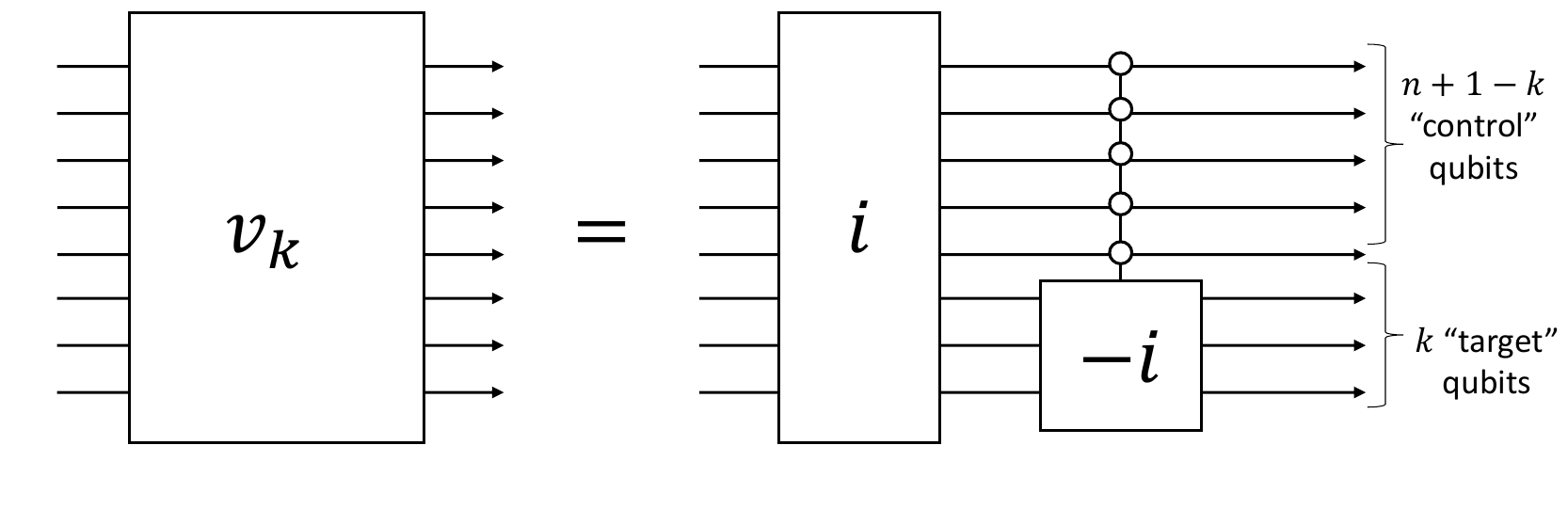}\caption{
			Quantum circuit diagram for the application of a relative phase $\iu$ to a specific subset of states. Even though all $n+1$ qubits comprise the target system in the overall schematic of Fig. \ref{fig:circuit diagram control}, we split them into $n+1-k$ qubits controlling the remaining $k$ target qubits for this schematic. When the first $n+1-k$ qubits are in state $\ket{g}$, a relative phase of $-\iu$ is applied. The overall phase of $\iu$ can be applied either before or after the relative phase.}\label{fig:circuit diagram V}
	\end{figure*}

	Next, we outline possible realizations of the unitaries $v_k$ from Eq. \eqref{eq:vk unitary}. These unitaries look like controlled gates, where a relative phase of $\iu$ is applied to every state whose first $n+1-k$ qubits are not all in state $\ket{g}$. We can rewrite these unitaries as
	\eq{
		v_k=\iu\left[\mathds{1}^{\otimes n+1}-\left(\ket{g}\bra{g}\right)^{\otimes n+1-k}\otimes\mathds{1}^{\otimes k}\right].
	} This is the application of an overall phase $\iu$ to the entire state and then a relative phase $-\iu$ controlled by the first $n+1-k$ qubits, which is depicted in Fig. \ref{fig:circuit diagram V}. Such multiply controlled gates can be further broken down into simpler gate models using the construction in Ref. \cite{SaeediPedram2013}.
	Overall, once HBAC has exhausted its usefulness, these steps provide a deterministic routine for cooling the vast majority of the qubits to their ground states using a single binary-outcome measurement that has arbitrarily large probability of success.
	
	\section{Concluding remarks} We have shown that protocols incorporating \sugg{coherent superpositions and postselection} can outperform HBAC protocols that were previously thought to be optimal. Our protocols make use of a quantum switch to apply \suggOLD{a superposition of} two unitary operations in place of the single unitary in standard algorithmic cooling protocols, deterministically generating a vast number of completely pure quantum states. 
	This solidifies the import of ICO in \suggOLD{\sugg{\textit{instructing}} us how to surpass} quantum limits that themselves already supersede classical bounds.
	
	A few final comments pertain. First, one must be aware of the cost of doing a regular HBAC protocol to generate the probability distribution $\boldsymbol{p}^\infty$; this must be repeated $m$ times to successfully generate the pure qubits. However, one can also calculate the effect that $\boldsymbol{T}_-$ has on $\boldsymbol{p}^\infty$, which may be negligible even when repeated a number of times, so one may only need to repeat the HBAC protocol a fraction of $m$ times while implementing \suggQ{controlled superpositions and postselection} $m$ times. \suggOLD{One may further select the optimal ICO protocol or ICO-inspired protocol whose $\boldsymbol{T}_-$ has the least effect on $\boldsymbol{p}^\infty$ \suggQ{and that requires the lowest value of $m$, such as HBAC+20PS with $m\approx 1$}.} 
	\suggOLD{Second, the HBAC part of the protocol can readily be replaced by the extended HBAC method developed in Ref. \cite{Alhambraetal2019} when it is physically convenient to diagonalize the target system in its energy eigenbasis and to 
		vary the dynamics between the thermal bath and the target system, so the cost of the HBAC portion of our protocol can thereby be diminished.} \suggQ{When those dynamics can be arbitrarily controlled, the system can always trivially be cooled; this is because adding an auxiliary system, performing an arbitrary joint interaction, then ignoring or measuring the arbitrary system allows one to enact an arbitrary quantum channel on a given quantum system to generate any desired state \cite{Wuetal2007} and strongly violates the assumptions of HBAC. The HBAC can also be replaced by any other procedure that helps to serially cool a quantum system toward its ground state, including other measurement-based protocols \cite{Eschneretal1995,Nakazatoetal2003,Nakazatoetal2004,Wuetal2004,Roaetal2006,Lietal2011,Machnesetal2012,Xuetal2014,Raoetal2016,Montenegroetal2018,Zhangetal2019,Pueblaetal2020,Yanetal2021,Konaretal2022,YanJing2022}.}
	\suggOLD{Third}, HBAC protocols have recently been shown to potentially benefit from the presence of environmental noise \cite{Farahmandetal2022}. This significantly improves the viability of all HBAC protocols, including the current proposal that augments HBAC with \sugg{postselection}. \suggOLD{Fourth}, if the control is slightly mixed, which can be evaded by a judicious choice of physical system for the control, the probability of creating $n$ pure qubits remains much higher than the corresponding probability using HBAC alone for realistic values of $\varepsilon$. \suggOLD{Fifth}, ICO has recently been linked to improvements in refrigeration \cite{FelceVedral2020,Nieetal2022inPress,Felceetal2021arxiv,Nieetal2022coolingexpt,Dieguezetal2023} and thermometry \cite{Mukhopadhyayetal2018arxiv}. Since purification is intimately linked to the amount of work that can be extracted from an ensemble of qubits, our purification protocol strongly supports the impact ICO may have on quantum thermodynamics.  Finally, these results are readily extendable to multiqudit systems 
	while still only requiring a two-level control system \sugg{and a single postselection measurement}.
	Taken together, we are confident that \suggQ{quantum switches and postselection} will continue to have a strong impact on quantum technologies including HBAC and beyond.

\begin{acknowledgments}
    The authors acknowledge that the NRC headquarters is located on the traditional unceded territory of the Algonquin Anishinaabe and Mohawk people. AZG acknowledges funding from the NSERC PDF program.
\end{acknowledgments}

\begin{appendix}
    \section{Calculating the transfer matrix}
    \label{app:transfer matrix}
    Begin with a $2^n$-component vector $\boldsymbol{p}$ containing the probability distribution for the $n$ system qubits in the computational basis. We have ignored the superscript $t$ from the main text that keeps track of the number of iterations of the HBAC protocol. Add a reset qubit to the system with density matrix given in Eq. \eqref{eq:reset qubit}. Since the reset qubit is also diagonal in the computational basis, standard tensor product rules dictate that the $n+1$ qubits be represented by a density matrix with $2^{n+1}$ diagonal components $\boldsymbol{\lambda}$. As well, since the overall state is separable between the system and the reset qubit, we must have that tracing out the system (reset qubit) yields the reset qubit (system). These together imply a specific relationship between the components of $\boldsymbol{p}$ and $\boldsymbol{\lambda}$ for all $1\leq k\leq 2^n$: 
    \eq{
    p_k&=\lambda^t_{2k-1}+\lambda^t_{2k}, 
    \\
    \lambda_{2k-1}&=p_k\frac{\eu^\varepsilon}{z},\\
    \lambda_{2k}&=p_k\frac{\eu^{-\varepsilon}}{z},
    } which is manifestly self-consistent because $z=2\cosh\varepsilon$. In matrix form, this looks like
    \eq{
    \boldsymbol{\lambda}=\begin{pmatrix}\boldsymbol{p}\frac{\eu^{\varepsilon}}{z}\\\boldsymbol{p}\frac{\eu^{-\varepsilon}}{z}
    \end{pmatrix}.
    } Next, transform the system together with the reset qubit according to Eqs. \eqref{eq:quantum switch} and \eqref{eq:U1 and U2 original}. We know that 
    \begin{equation}
	\begin{aligned}
		U_AU_B&=(U_BU_A)^\dagger=\mathrm{Diag}(
		1,
		\underbrace{\iu\sigma_x,\cdots,\iu\sigma_x}_{2^n-1\,\mathrm{times}},1
		). 
		\end{aligned}
	\end{equation} Measuring the control to be in $\ket{+}$ implies using the $+$ parts of the $\pm$ symbol in Eq. \eqref{eq:quantum switch} to enact
	\eq{
	\rho\to (U+U^\dagger)\rho(U+U^\dagger)/4P_+
	} for unitary $U=U_BU_A$. The inteference terms lead to most of the components canceling in
	\eq{
	U_+\equiv U+U^\dagger=\mathrm{Diag}(
		2,
		\underbrace{0,\cdots,0}_{2^{n+1}-2\,\mathrm{times}},2
		). 
	} It would not be possible to simply enact $U_+$  because it is manifestly nonunitary; this is the ability conferred by controlled superpositions and postselection.
	Then, since the unitaries and the state are diagonal in the computational basis, we directly evaluate
	\begin{widetext}
	\eq{
	    \frac{1}{4P_+}\begin{pmatrix}
	        2&0&\cdots&0&0\\
	        0&0&\cdots&0&0\\
	        \vdots&\vdots&\ddots&\vdots&\vdots\\
	        0&0&\cdots&0&0\\
	        0&0&\cdots&0&2
	    \end{pmatrix}\begin{pmatrix}
	        \lambda_1&0&\cdots&0&0\\
	        0&\lambda_2&\cdots&0&0\\
	        \vdots&\vdots&\ddots&\vdots&\vdots\\
	        0&0&\cdots&\lambda_{2^{n+1}-1}&0\\
	        0&0&\cdots&0&\lambda_{2^{n+1}}
	    \end{pmatrix}\begin{pmatrix}
	        2&0&\cdots&0&0\\
	        0&0&\cdots&0&0\\
	        \vdots&\vdots&\ddots&\vdots&\vdots\\
	        0&0&\cdots&0&0\\
	        0&0&\cdots&0&2
	    \end{pmatrix}=
	    \frac{1}{P_+}\begin{pmatrix}
	        \lambda_1&0&\cdots&0&0\\
	        0&0&\cdots&0&0\\
	        \vdots&\vdots&\ddots&\vdots&\vdots\\
	        0&0&\cdots&0&0\\
	        0&0&\cdots&0&\lambda_{2^{n+1}}
	    \end{pmatrix}=\frac{1}{P_+}\begin{pmatrix}
	        p_1\frac{\eu^{\varepsilon}}{z}&0&\cdots&0&0\\
	        0&0&\cdots&0&0\\
	        \vdots&\vdots&\ddots&\vdots&\vdots\\
	        0&0&\cdots&0&0\\
	        0&0&\cdots&0&p_{2^n}\frac{\eu^{-\varepsilon}}{z}
	    \end{pmatrix}.
	}
	\end{widetext} The transfer matrix for $\boldsymbol{\lambda}\to \boldsymbol{S}_+\boldsymbol{\lambda}$ can be read off as
	\eq{
	\boldsymbol{S}_+\propto \mathrm{Diag}(
		1,
		\underbrace{0,\cdots,0}_{2^{n+1}-2\,\mathrm{times}},1
		).
	} As for the evolution of the system itself, we trace over the reset qubit to observe the overall evolution \eq{\boldsymbol{p}\to \frac{1}{P_+}(
	p_1\frac{\eu^{\varepsilon}}{z},
		\underbrace{0,\cdots,0}_{2^{n}-2\,\mathrm{times}},p_{2^n}\frac{\eu^{-\varepsilon}}{z}
		)^\top.} This yields the transfer matrix $\boldsymbol{T}_+$ provided in the main text.
    \end{appendix}


\begin{thebibliography}{90}%
	\makeatletter
	\providecommand \@ifxundefined [1]{%
		\@ifx{#1\undefined}
	}%
	\providecommand \@ifnum [1]{%
		\ifnum #1\expandafter \@firstoftwo
		\else \expandafter \@secondoftwo
		\fi
	}%
	\providecommand \@ifx [1]{%
		\ifx #1\expandafter \@firstoftwo
		\else \expandafter \@secondoftwo
		\fi
	}%
	\providecommand \natexlab [1]{#1}%
	\providecommand \enquote  [1]{``#1''}%
	\providecommand \bibnamefont  [1]{#1}%
	\providecommand \bibfnamefont [1]{#1}%
	\providecommand \citenamefont [1]{#1}%
	\providecommand \href@noop [0]{\@secondoftwo}%
	\providecommand \href [0]{\begingroup \@sanitize@url \@href}%
	\providecommand \@href[1]{\@@startlink{#1}\@@href}%
	\providecommand \@@href[1]{\endgroup#1\@@endlink}%
	\providecommand \@sanitize@url [0]{\catcode `\\12\catcode `\$12\catcode
		`\&12\catcode `\#12\catcode `\^12\catcode `\_12\catcode `\%12\relax}%
	\providecommand \@@startlink[1]{}%
	\providecommand \@@endlink[0]{}%
	\providecommand \url  [0]{\begingroup\@sanitize@url \@url }%
	\providecommand \@url [1]{\endgroup\@href {#1}{\urlprefix }}%
	\providecommand \urlprefix  [0]{URL }%
	\providecommand \Eprint [0]{\href }%
	\providecommand \doibase [0]{https://doi.org/}%
	\providecommand \selectlanguage [0]{\@gobble}%
	\providecommand \bibinfo  [0]{\@secondoftwo}%
	\providecommand \bibfield  [0]{\@secondoftwo}%
	\providecommand \translation [1]{[#1]}%
	\providecommand \BibitemOpen [0]{}%
	\providecommand \bibitemStop [0]{}%
	\providecommand \bibitemNoStop [0]{.\EOS\space}%
	\providecommand \EOS [0]{\spacefactor3000\relax}%
	\providecommand \BibitemShut  [1]{\csname bibitem#1\endcsname}%
	\let\auto@bib@innerbib\@empty
	\bibitem [{\citenamefont {Schulman}\ and\ \citenamefont
		{Vazirani}(1999)}]{Schulmanetal1999}%
	\BibitemOpen
	\bibfield  {author} {\bibinfo {author} {\bibfnamefont {L.~J.}\ \bibnamefont
			{Schulman}}\ and\ \bibinfo {author} {\bibfnamefont {U.~V.}\ \bibnamefont
			{Vazirani}},\ }\bibfield  {title} {\bibinfo {title} {Molecular scale heat
			engines and scalable quantum computation},\ }in\ \href
	{https://doi.org/10.1145/301250.301332} {\emph {\bibinfo {booktitle}
			{Proceedings of the Thirty-First Annual ACM Symposium on Theory of
				Computing}}},\ \bibinfo {series and number} {STOC '99}\ (\bibinfo
	{publisher} {Association for Computing Machinery},\ \bibinfo {address} {New
		York, NY, USA},\ \bibinfo {year} {1999})\ p.\ \bibinfo {pages}
	{322–329}\BibitemShut {NoStop}%
	\bibitem [{\citenamefont {Boykin}\ \emph {et~al.}(2002)\citenamefont {Boykin},
		\citenamefont {Mor}, \citenamefont {Roychowdhury}, \citenamefont {Vatan},\
		and\ \citenamefont {Vrijen}}]{Boykinetal2002}%
	\BibitemOpen
	\bibfield  {author} {\bibinfo {author} {\bibfnamefont {P.~O.}\ \bibnamefont
			{Boykin}}, \bibinfo {author} {\bibfnamefont {T.}~\bibnamefont {Mor}},
		\bibinfo {author} {\bibfnamefont {V.}~\bibnamefont {Roychowdhury}}, \bibinfo
		{author} {\bibfnamefont {F.}~\bibnamefont {Vatan}},\ and\ \bibinfo {author}
		{\bibfnamefont {R.}~\bibnamefont {Vrijen}},\ }\bibfield  {title} {\bibinfo
		{title} {Algorithmic cooling and scalable {NMR} quantum computers},\ }\href
	{https://doi.org/10.1073/pnas.241641898} {\bibfield  {journal} {\bibinfo
			{journal} {Proceedings of the National Academy of Sciences}\ }\textbf
		{\bibinfo {volume} {99}},\ \bibinfo {pages} {3388} (\bibinfo {year}
		{2002})}\BibitemShut {NoStop}%
	\bibitem [{\citenamefont {Fernandez}\ \emph {et~al.}(2004)\citenamefont
		{Fernandez}, \citenamefont {Lloyd}, \citenamefont {Mor},\ and\ \citenamefont
		{Roychowdhury}}]{Fernandezetal2004}%
	\BibitemOpen
	\bibfield  {author} {\bibinfo {author} {\bibfnamefont {J.~M.}\ \bibnamefont
			{Fernandez}}, \bibinfo {author} {\bibfnamefont {S.}~\bibnamefont {Lloyd}},
		\bibinfo {author} {\bibfnamefont {T.}~\bibnamefont {Mor}},\ and\ \bibinfo
		{author} {\bibfnamefont {V.}~\bibnamefont {Roychowdhury}},\ }\bibfield
	{title} {\bibinfo {title} {Algorithmic cooling of spins: A practicable method
			for increasing polarization},\ }\href
	{https://doi.org/10.1142/S0219749904000419} {\bibfield  {journal} {\bibinfo
			{journal} {International Journal of Quantum Information}\ }\textbf {\bibinfo
			{volume} {02}},\ \bibinfo {pages} {461} (\bibinfo {year} {2004})}\BibitemShut
	{NoStop}%
	\bibitem [{\citenamefont {Schulman}\ \emph {et~al.}(2005)\citenamefont
		{Schulman}, \citenamefont {Mor},\ and\ \citenamefont
		{Weinstein}}]{Schulmanetal2005}%
	\BibitemOpen
	\bibfield  {author} {\bibinfo {author} {\bibfnamefont {L.~J.}\ \bibnamefont
			{Schulman}}, \bibinfo {author} {\bibfnamefont {T.}~\bibnamefont {Mor}},\ and\
		\bibinfo {author} {\bibfnamefont {Y.}~\bibnamefont {Weinstein}},\ }\bibfield
	{title} {\bibinfo {title} {Physical limits of heat-bath algorithmic
			cooling},\ }\href {https://doi.org/10.1103/PhysRevLett.94.120501} {\bibfield
		{journal} {\bibinfo  {journal} {Phys. Rev. Lett.}\ }\textbf {\bibinfo
			{volume} {94}},\ \bibinfo {pages} {120501} (\bibinfo {year}
		{2005})}\BibitemShut {NoStop}%
	\bibitem [{\citenamefont {Raeisi}\ and\ \citenamefont
		{Mosca}(2015)}]{RaeisiMosca2015}%
	\BibitemOpen
	\bibfield  {author} {\bibinfo {author} {\bibfnamefont {S.}~\bibnamefont
			{Raeisi}}\ and\ \bibinfo {author} {\bibfnamefont {M.}~\bibnamefont {Mosca}},\
	}\bibfield  {title} {\bibinfo {title} {Asymptotic bound for heat-bath
			algorithmic cooling},\ }\href
	{https://doi.org/10.1103/PhysRevLett.114.100404} {\bibfield  {journal}
		{\bibinfo  {journal} {Phys. Rev. Lett.}\ }\textbf {\bibinfo {volume} {114}},\
		\bibinfo {pages} {100404} (\bibinfo {year} {2015})}\BibitemShut {NoStop}%
	\bibitem [{\citenamefont {Rodr\'{\i}guez-Briones}\ and\ \citenamefont
		{Laflamme}(2016)}]{RodriguezBrionesetal2016}%
	\BibitemOpen
	\bibfield  {author} {\bibinfo {author} {\bibfnamefont {N.~A.}\ \bibnamefont
			{Rodr\'{\i}guez-Briones}}\ and\ \bibinfo {author} {\bibfnamefont
			{R.}~\bibnamefont {Laflamme}},\ }\bibfield  {title} {\bibinfo {title}
		{Achievable polarization for heat-bath algorithmic cooling},\ }\href
	{https://doi.org/10.1103/PhysRevLett.116.170501} {\bibfield  {journal}
		{\bibinfo  {journal} {Phys. Rev. Lett.}\ }\textbf {\bibinfo {volume} {116}},\
		\bibinfo {pages} {170501} (\bibinfo {year} {2016})}\BibitemShut {NoStop}%
	\bibitem [{\citenamefont {Rodr{\'{\i}}guez-Briones}\ \emph
		{et~al.}(2017)\citenamefont {Rodr{\'{\i}}guez-Briones}, \citenamefont {Li},
		\citenamefont {Peng}, \citenamefont {Mor}, \citenamefont {Weinstein},\ and\
		\citenamefont {Laflamme}}]{RodriguezBriones2017}%
	\BibitemOpen
	\bibfield  {author} {\bibinfo {author} {\bibfnamefont {N.~A.}\ \bibnamefont
			{Rodr{\'{\i}}guez-Briones}}, \bibinfo {author} {\bibfnamefont
			{J.}~\bibnamefont {Li}}, \bibinfo {author} {\bibfnamefont {X.}~\bibnamefont
			{Peng}}, \bibinfo {author} {\bibfnamefont {T.}~\bibnamefont {Mor}}, \bibinfo
		{author} {\bibfnamefont {Y.}~\bibnamefont {Weinstein}},\ and\ \bibinfo
		{author} {\bibfnamefont {R.}~\bibnamefont {Laflamme}},\ }\bibfield  {title}
	{\bibinfo {title} {Heat-bath algorithmic cooling with correlated
			qubit-environment interactions},\ }\href
	{https://doi.org/10.1088/1367-2630/aa8fe0} {\bibfield  {journal} {\bibinfo
			{journal} {New Journal of Physics}\ }\textbf {\bibinfo {volume} {19}},\
		\bibinfo {pages} {113047} (\bibinfo {year} {2017})}\BibitemShut {NoStop}%
	\bibitem [{\citenamefont {Raeisi}\ \emph {et~al.}(2019)\citenamefont {Raeisi},
		\citenamefont {Kieferov\'a},\ and\ \citenamefont {Mosca}}]{Raeisietal2019}%
	\BibitemOpen
	\bibfield  {author} {\bibinfo {author} {\bibfnamefont {S.}~\bibnamefont
			{Raeisi}}, \bibinfo {author} {\bibfnamefont {M.}~\bibnamefont
			{Kieferov\'a}},\ and\ \bibinfo {author} {\bibfnamefont {M.}~\bibnamefont
			{Mosca}},\ }\bibfield  {title} {\bibinfo {title} {Novel technique for robust
			optimal algorithmic cooling},\ }\href
	{https://doi.org/10.1103/PhysRevLett.122.220501} {\bibfield  {journal}
		{\bibinfo  {journal} {Phys. Rev. Lett.}\ }\textbf {\bibinfo {volume} {122}},\
		\bibinfo {pages} {220501} (\bibinfo {year} {2019})}\BibitemShut {NoStop}%
	\bibitem [{\citenamefont {Alhambra}\ \emph {et~al.}(2019)\citenamefont
		{Alhambra}, \citenamefont {Lostaglio},\ and\ \citenamefont
		{Perry}}]{Alhambraetal2019}%
	\BibitemOpen
	\bibfield  {author} {\bibinfo {author} {\bibfnamefont {{\'{A}}.~M.}\
			\bibnamefont {Alhambra}}, \bibinfo {author} {\bibfnamefont {M.}~\bibnamefont
			{Lostaglio}},\ and\ \bibinfo {author} {\bibfnamefont {C.}~\bibnamefont
			{Perry}},\ }\bibfield  {title} {\bibinfo {title} {Heat-{B}ath {A}lgorithmic
			{C}ooling with optimal thermalization strategies},\ }\href
	{https://doi.org/10.22331/q-2019-09-23-188} {\bibfield  {journal} {\bibinfo
			{journal} {{Quantum}}\ }\textbf {\bibinfo {volume} {3}},\ \bibinfo {pages}
		{188} (\bibinfo {year} {2019})}\BibitemShut {NoStop}%
	\bibitem [{\citenamefont {{Pande}}(2020)}]{Pande2020arxiv}%
	\BibitemOpen
	\bibfield  {author} {\bibinfo {author} {\bibfnamefont {V.~R.}\ \bibnamefont
			{{Pande}}},\ }\bibfield  {title} {\bibinfo {title} {{Optimal Entropy
				Compression and Purification in Quantum Bits}},\ }\href@noop {} {\  (\bibinfo
		{year} {2020})},\ \Eprint {https://arxiv.org/abs/2001.00562}
	{arXiv:2001.00562 [quant-ph]} \BibitemShut {NoStop}%
	\bibitem [{\citenamefont {Raeisi}(2021)}]{Raeisi2021}%
	\BibitemOpen
	\bibfield  {author} {\bibinfo {author} {\bibfnamefont {S.}~\bibnamefont
			{Raeisi}},\ }\bibfield  {title} {\bibinfo {title} {No-go theorem behind the
			limit of the heat-bath algorithmic cooling},\ }\href
	{https://doi.org/10.1103/PhysRevA.103.062424} {\bibfield  {journal} {\bibinfo
			{journal} {Phys. Rev. A}\ }\textbf {\bibinfo {volume} {103}},\ \bibinfo
		{pages} {062424} (\bibinfo {year} {2021})}\BibitemShut {NoStop}%
	\bibitem [{\citenamefont {Farahmand}\ \emph {et~al.}(2022)\citenamefont
		{Farahmand}, \citenamefont {Aghaei~Saem},\ and\ \citenamefont
		{Raeisi}}]{Farahmandetal2022}%
	\BibitemOpen
	\bibfield  {author} {\bibinfo {author} {\bibfnamefont {Z.}~\bibnamefont
			{Farahmand}}, \bibinfo {author} {\bibfnamefont {R.}~\bibnamefont
			{Aghaei~Saem}},\ and\ \bibinfo {author} {\bibfnamefont {S.}~\bibnamefont
			{Raeisi}},\ }\bibfield  {title} {\bibinfo {title} {Quantum noise can enhance
			algorithmic cooling},\ }\href {https://doi.org/10.1103/PhysRevA.105.022418}
	{\bibfield  {journal} {\bibinfo  {journal} {Phys. Rev. A}\ }\textbf {\bibinfo
			{volume} {105}},\ \bibinfo {pages} {022418} (\bibinfo {year}
		{2022})}\BibitemShut {NoStop}%
	\bibitem [{\citenamefont {Baugh}\ \emph {et~al.}(2005)\citenamefont {Baugh},
		\citenamefont {Moussa}, \citenamefont {Ryan}, \citenamefont {Nayak},\ and\
		\citenamefont {Laflamme}}]{Baughetal2005}%
	\BibitemOpen
	\bibfield  {author} {\bibinfo {author} {\bibfnamefont {J.}~\bibnamefont
			{Baugh}}, \bibinfo {author} {\bibfnamefont {O.}~\bibnamefont {Moussa}},
		\bibinfo {author} {\bibfnamefont {C.~A.}\ \bibnamefont {Ryan}}, \bibinfo
		{author} {\bibfnamefont {A.}~\bibnamefont {Nayak}},\ and\ \bibinfo {author}
		{\bibfnamefont {R.}~\bibnamefont {Laflamme}},\ }\bibfield  {title} {\bibinfo
		{title} {Experimental implementation of heat-bath algorithmic cooling using
			solid-state nuclear magnetic resonance},\ }\href
	{https://doi.org/10.1038/nature04272} {\bibfield  {journal} {\bibinfo
			{journal} {Nature}\ }\textbf {\bibinfo {volume} {438}},\ \bibinfo {pages}
		{470} (\bibinfo {year} {2005})}\BibitemShut {NoStop}%
	\bibitem [{\citenamefont {{Zaiser}}\ \emph {et~al.}(2018)\citenamefont
		{{Zaiser}}, \citenamefont {{Masatth}}, \citenamefont {{Bhaktavatsala Rao}},
		\citenamefont {{Raeisi}},\ and\ \citenamefont
		{{Wrachtrup}}}]{Zaiseretal2018arxiv}%
	\BibitemOpen
	\bibfield  {author} {\bibinfo {author} {\bibfnamefont {S.}~\bibnamefont
			{{Zaiser}}}, \bibinfo {author} {\bibfnamefont {B.}~\bibnamefont {{Masatth}}},
		\bibinfo {author} {\bibfnamefont {D.~D.}\ \bibnamefont {{Bhaktavatsala
					Rao}}}, \bibinfo {author} {\bibfnamefont {S.}~\bibnamefont {{Raeisi}}},\ and\
		\bibinfo {author} {\bibfnamefont {J.}~\bibnamefont {{Wrachtrup}}},\
	}\bibfield  {title} {\bibinfo {title} {{Experimental Saturation of the
				Heat-Bath Algorithmic Cooling bound}},\ }\href@noop {} {\  (\bibinfo {year}
		{2018})},\ \Eprint {https://arxiv.org/abs/1812.06252} {arXiv:1812.06252
		[quant-ph]} \BibitemShut {NoStop}%
	\bibitem [{Note1()}]{Note1}%
	\BibitemOpen
	\bibinfo {note} {If one assumes that the thermal bath is a resource whose
		dynamics can be manipulated, that one has access to \protect \textit {a
			priori} knowledge of the states to be purified, and that the thermal bath is
		sufficiently cold, the extended HBAC methods presented in Ref. \cite
		{Alhambraetal2019} can increase the probability of creating a purified
		state.}\BibitemShut {Stop}%
	\bibitem [{\citenamefont {Knill}\ and\ \citenamefont
		{Laflamme}(1998)}]{KnillLaflamme1998}%
	\BibitemOpen
	\bibfield  {author} {\bibinfo {author} {\bibfnamefont {E.}~\bibnamefont
			{Knill}}\ and\ \bibinfo {author} {\bibfnamefont {R.}~\bibnamefont
			{Laflamme}},\ }\bibfield  {title} {\bibinfo {title} {Power of one bit of
			quantum information},\ }\href {https://doi.org/10.1103/PhysRevLett.81.5672}
	{\bibfield  {journal} {\bibinfo  {journal} {Phys. Rev. Lett.}\ }\textbf
		{\bibinfo {volume} {81}},\ \bibinfo {pages} {5672} (\bibinfo {year}
		{1998})}\BibitemShut {NoStop}%
	\bibitem [{\citenamefont {Eschner}\ \emph {et~al.}(1995)\citenamefont
		{Eschner}, \citenamefont {Appasamy},\ and\ \citenamefont
		{Toschek}}]{Eschneretal1995}%
	\BibitemOpen
	\bibfield  {author} {\bibinfo {author} {\bibfnamefont {J.}~\bibnamefont
			{Eschner}}, \bibinfo {author} {\bibfnamefont {B.}~\bibnamefont {Appasamy}},\
		and\ \bibinfo {author} {\bibfnamefont {P.~E.}\ \bibnamefont {Toschek}},\
	}\bibfield  {title} {\bibinfo {title} {Stochastic cooling of a trapped ion by
			null detection of its fluorescence},\ }\href
	{https://doi.org/10.1103/PhysRevLett.74.2435} {\bibfield  {journal} {\bibinfo
			{journal} {Phys. Rev. Lett.}\ }\textbf {\bibinfo {volume} {74}},\ \bibinfo
		{pages} {2435} (\bibinfo {year} {1995})}\BibitemShut {NoStop}%
	\bibitem [{\citenamefont {Nakazato}\ \emph {et~al.}(2003)\citenamefont
		{Nakazato}, \citenamefont {Takazawa},\ and\ \citenamefont
		{Yuasa}}]{Nakazatoetal2003}%
	\BibitemOpen
	\bibfield  {author} {\bibinfo {author} {\bibfnamefont {H.}~\bibnamefont
			{Nakazato}}, \bibinfo {author} {\bibfnamefont {T.}~\bibnamefont {Takazawa}},\
		and\ \bibinfo {author} {\bibfnamefont {K.}~\bibnamefont {Yuasa}},\ }\bibfield
	{title} {\bibinfo {title} {Purification through zeno-like measurements},\
	}\href {https://doi.org/10.1103/PhysRevLett.90.060401} {\bibfield  {journal}
		{\bibinfo  {journal} {Phys. Rev. Lett.}\ }\textbf {\bibinfo {volume} {90}},\
		\bibinfo {pages} {060401} (\bibinfo {year} {2003})}\BibitemShut {NoStop}%
	\bibitem [{\citenamefont {Nakazato}\ \emph {et~al.}(2004)\citenamefont
		{Nakazato}, \citenamefont {Unoki},\ and\ \citenamefont
		{Yuasa}}]{Nakazatoetal2004}%
	\BibitemOpen
	\bibfield  {author} {\bibinfo {author} {\bibfnamefont {H.}~\bibnamefont
			{Nakazato}}, \bibinfo {author} {\bibfnamefont {M.}~\bibnamefont {Unoki}},\
		and\ \bibinfo {author} {\bibfnamefont {K.}~\bibnamefont {Yuasa}},\ }\bibfield
	{title} {\bibinfo {title} {Preparation and entanglement purification of
			qubits through zeno-like measurements},\ }\href
	{https://doi.org/10.1103/PhysRevA.70.012303} {\bibfield  {journal} {\bibinfo
			{journal} {Phys. Rev. A}\ }\textbf {\bibinfo {volume} {70}},\ \bibinfo
		{pages} {012303} (\bibinfo {year} {2004})}\BibitemShut {NoStop}%
	\bibitem [{\citenamefont {Wu}\ \emph {et~al.}(2004)\citenamefont {Wu},
		\citenamefont {Lidar},\ and\ \citenamefont {Schneider}}]{Wuetal2004}%
	\BibitemOpen
	\bibfield  {author} {\bibinfo {author} {\bibfnamefont {L.-A.}\ \bibnamefont
			{Wu}}, \bibinfo {author} {\bibfnamefont {D.~A.}\ \bibnamefont {Lidar}},\ and\
		\bibinfo {author} {\bibfnamefont {S.}~\bibnamefont {Schneider}},\ }\bibfield
	{title} {\bibinfo {title} {Long-range entanglement generation via frequent
			measurements},\ }\href {https://doi.org/10.1103/PhysRevA.70.032322}
	{\bibfield  {journal} {\bibinfo  {journal} {Phys. Rev. A}\ }\textbf {\bibinfo
			{volume} {70}},\ \bibinfo {pages} {032322} (\bibinfo {year}
		{2004})}\BibitemShut {NoStop}%
	\bibitem [{\citenamefont {Roa}\ \emph {et~al.}(2006)\citenamefont {Roa},
		\citenamefont {Delgado}, \citenamefont {Ladr\'on~de Guevara},\ and\
		\citenamefont {Klimov}}]{Roaetal2006}%
	\BibitemOpen
	\bibfield  {author} {\bibinfo {author} {\bibfnamefont {L.}~\bibnamefont
			{Roa}}, \bibinfo {author} {\bibfnamefont {A.}~\bibnamefont {Delgado}},
		\bibinfo {author} {\bibfnamefont {M.~L.}\ \bibnamefont {Ladr\'on~de
				Guevara}},\ and\ \bibinfo {author} {\bibfnamefont {A.~B.}\ \bibnamefont
			{Klimov}},\ }\bibfield  {title} {\bibinfo {title} {Measurement-driven quantum
			evolution},\ }\href {https://doi.org/10.1103/PhysRevA.73.012322} {\bibfield
		{journal} {\bibinfo  {journal} {Phys. Rev. A}\ }\textbf {\bibinfo {volume}
			{73}},\ \bibinfo {pages} {012322} (\bibinfo {year} {2006})}\BibitemShut
	{NoStop}%
	\bibitem [{\citenamefont {Li}\ \emph {et~al.}(2011)\citenamefont {Li},
		\citenamefont {Wu}, \citenamefont {Wang},\ and\ \citenamefont
		{Yang}}]{Lietal2011}%
	\BibitemOpen
	\bibfield  {author} {\bibinfo {author} {\bibfnamefont {Y.}~\bibnamefont
			{Li}}, \bibinfo {author} {\bibfnamefont {L.-A.}\ \bibnamefont {Wu}}, \bibinfo
		{author} {\bibfnamefont {Y.-D.}\ \bibnamefont {Wang}},\ and\ \bibinfo
		{author} {\bibfnamefont {L.-P.}\ \bibnamefont {Yang}},\ }\bibfield  {title}
	{\bibinfo {title} {Nondeterministic ultrafast ground-state cooling of a
			mechanical resonator},\ }\href {https://doi.org/10.1103/PhysRevB.84.094502}
	{\bibfield  {journal} {\bibinfo  {journal} {Phys. Rev. B}\ }\textbf {\bibinfo
			{volume} {84}},\ \bibinfo {pages} {094502} (\bibinfo {year}
		{2011})}\BibitemShut {NoStop}%
	\bibitem [{\citenamefont {Machnes}\ \emph {et~al.}(2012)\citenamefont
		{Machnes}, \citenamefont {Cerrillo}, \citenamefont {Aspelmeyer},
		\citenamefont {Wieczorek}, \citenamefont {Plenio},\ and\ \citenamefont
		{Retzker}}]{Machnesetal2012}%
	\BibitemOpen
	\bibfield  {author} {\bibinfo {author} {\bibfnamefont {S.}~\bibnamefont
			{Machnes}}, \bibinfo {author} {\bibfnamefont {J.}~\bibnamefont {Cerrillo}},
		\bibinfo {author} {\bibfnamefont {M.}~\bibnamefont {Aspelmeyer}}, \bibinfo
		{author} {\bibfnamefont {W.}~\bibnamefont {Wieczorek}}, \bibinfo {author}
		{\bibfnamefont {M.~B.}\ \bibnamefont {Plenio}},\ and\ \bibinfo {author}
		{\bibfnamefont {A.}~\bibnamefont {Retzker}},\ }\bibfield  {title} {\bibinfo
		{title} {Pulsed laser cooling for cavity optomechanical resonators},\ }\href
	{https://doi.org/10.1103/PhysRevLett.108.153601} {\bibfield  {journal}
		{\bibinfo  {journal} {Phys. Rev. Lett.}\ }\textbf {\bibinfo {volume} {108}},\
		\bibinfo {pages} {153601} (\bibinfo {year} {2012})}\BibitemShut {NoStop}%
	\bibitem [{\citenamefont {Xu}\ \emph {et~al.}(2014)\citenamefont {Xu},
		\citenamefont {Yung}, \citenamefont {Xu}, \citenamefont {Boixo},
		\citenamefont {Zhou}, \citenamefont {Li}, \citenamefont {Aspuru-Guzik},\ and\
		\citenamefont {Guo}}]{Xuetal2014}%
	\BibitemOpen
	\bibfield  {author} {\bibinfo {author} {\bibfnamefont {J.-S.}\ \bibnamefont
			{Xu}}, \bibinfo {author} {\bibfnamefont {M.-H.}\ \bibnamefont {Yung}},
		\bibinfo {author} {\bibfnamefont {X.-Y.}\ \bibnamefont {Xu}}, \bibinfo
		{author} {\bibfnamefont {S.}~\bibnamefont {Boixo}}, \bibinfo {author}
		{\bibfnamefont {Z.-W.}\ \bibnamefont {Zhou}}, \bibinfo {author}
		{\bibfnamefont {C.-F.}\ \bibnamefont {Li}}, \bibinfo {author} {\bibfnamefont
			{A.}~\bibnamefont {Aspuru-Guzik}},\ and\ \bibinfo {author} {\bibfnamefont
			{G.-C.}\ \bibnamefont {Guo}},\ }\bibfield  {title} {\bibinfo {title}
		{Demon-like algorithmic quantum cooling and its realization with quantum
			optics},\ }\href {https://doi.org/10.1038/nphoton.2013.354} {\bibfield
		{journal} {\bibinfo  {journal} {Nature Photonics}\ }\textbf {\bibinfo
			{volume} {8}},\ \bibinfo {pages} {113} (\bibinfo {year} {2014})}\BibitemShut
	{NoStop}%
	\bibitem [{\citenamefont {Rao}\ \emph {et~al.}(2016)\citenamefont {Rao},
		\citenamefont {Momenzadeh},\ and\ \citenamefont {Wrachtrup}}]{Raoetal2016}%
	\BibitemOpen
	\bibfield  {author} {\bibinfo {author} {\bibfnamefont {D.~D.~B.}\
			\bibnamefont {Rao}}, \bibinfo {author} {\bibfnamefont {S.~A.}\ \bibnamefont
			{Momenzadeh}},\ and\ \bibinfo {author} {\bibfnamefont {J.}~\bibnamefont
			{Wrachtrup}},\ }\bibfield  {title} {\bibinfo {title} {Heralded control of
			mechanical motion by single spins},\ }\href
	{https://doi.org/10.1103/PhysRevLett.117.077203} {\bibfield  {journal}
		{\bibinfo  {journal} {Phys. Rev. Lett.}\ }\textbf {\bibinfo {volume} {117}},\
		\bibinfo {pages} {077203} (\bibinfo {year} {2016})}\BibitemShut {NoStop}%
	\bibitem [{\citenamefont {Montenegro}\ \emph {et~al.}(2018)\citenamefont
		{Montenegro}, \citenamefont {Coto}, \citenamefont {Eremeev},\ and\
		\citenamefont {Orszag}}]{Montenegroetal2018}%
	\BibitemOpen
	\bibfield  {author} {\bibinfo {author} {\bibfnamefont {V.}~\bibnamefont
			{Montenegro}}, \bibinfo {author} {\bibfnamefont {R.}~\bibnamefont {Coto}},
		\bibinfo {author} {\bibfnamefont {V.}~\bibnamefont {Eremeev}},\ and\ \bibinfo
		{author} {\bibfnamefont {M.}~\bibnamefont {Orszag}},\ }\bibfield  {title}
	{\bibinfo {title} {Ground-state cooling of a nanomechanical oscillator with
			$n$ spins},\ }\href {https://doi.org/10.1103/PhysRevA.98.053837} {\bibfield
		{journal} {\bibinfo  {journal} {Phys. Rev. A}\ }\textbf {\bibinfo {volume}
			{98}},\ \bibinfo {pages} {053837} (\bibinfo {year} {2018})}\BibitemShut
	{NoStop}%
	\bibitem [{\citenamefont {Zhang}\ \emph {et~al.}(2019)\citenamefont {Zhang},
		\citenamefont {Jing}, \citenamefont {Wu}, \citenamefont {Wang},\ and\
		\citenamefont {Zhu}}]{Zhangetal2019}%
	\BibitemOpen
	\bibfield  {author} {\bibinfo {author} {\bibfnamefont {J.-M.}\ \bibnamefont
			{Zhang}}, \bibinfo {author} {\bibfnamefont {J.}~\bibnamefont {Jing}},
		\bibinfo {author} {\bibfnamefont {L.-A.}\ \bibnamefont {Wu}}, \bibinfo
		{author} {\bibfnamefont {L.-G.}\ \bibnamefont {Wang}},\ and\ \bibinfo
		{author} {\bibfnamefont {S.-Y.}\ \bibnamefont {Zhu}},\ }\bibfield  {title}
	{\bibinfo {title} {Measurement-induced cooling of a qubit in structured
			environments},\ }\href {https://doi.org/10.1103/PhysRevA.100.022107}
	{\bibfield  {journal} {\bibinfo  {journal} {Phys. Rev. A}\ }\textbf {\bibinfo
			{volume} {100}},\ \bibinfo {pages} {022107} (\bibinfo {year}
		{2019})}\BibitemShut {NoStop}%
	\bibitem [{\citenamefont {Puebla}\ \emph {et~al.}(2020)\citenamefont {Puebla},
		\citenamefont {Abah},\ and\ \citenamefont {Paternostro}}]{Pueblaetal2020}%
	\BibitemOpen
	\bibfield  {author} {\bibinfo {author} {\bibfnamefont {R.}~\bibnamefont
			{Puebla}}, \bibinfo {author} {\bibfnamefont {O.}~\bibnamefont {Abah}},\ and\
		\bibinfo {author} {\bibfnamefont {M.}~\bibnamefont {Paternostro}},\
	}\bibfield  {title} {\bibinfo {title} {Measurement-based cooling of a
			nonlinear mechanical resonator},\ }\href
	{https://doi.org/10.1103/PhysRevB.101.245410} {\bibfield  {journal} {\bibinfo
			{journal} {Phys. Rev. B}\ }\textbf {\bibinfo {volume} {101}},\ \bibinfo
		{pages} {245410} (\bibinfo {year} {2020})}\BibitemShut {NoStop}%
	\bibitem [{\citenamefont {Yan}\ and\ \citenamefont {Jing}(2021)}]{Yanetal2021}%
	\BibitemOpen
	\bibfield  {author} {\bibinfo {author} {\bibfnamefont {J.-s.}\ \bibnamefont
			{Yan}}\ and\ \bibinfo {author} {\bibfnamefont {J.}~\bibnamefont {Jing}},\
	}\bibfield  {title} {\bibinfo {title} {External-level assisted cooling by
			measurement},\ }\href {https://doi.org/10.1103/PhysRevA.104.063105}
	{\bibfield  {journal} {\bibinfo  {journal} {Phys. Rev. A}\ }\textbf {\bibinfo
			{volume} {104}},\ \bibinfo {pages} {063105} (\bibinfo {year}
		{2021})}\BibitemShut {NoStop}%
	\bibitem [{\citenamefont {Konar}\ \emph {et~al.}(2022)\citenamefont {Konar},
		\citenamefont {Ghosh},\ and\ \citenamefont {Sen(De)}}]{Konaretal2022}%
	\BibitemOpen
	\bibfield  {author} {\bibinfo {author} {\bibfnamefont {T.~K.}\ \bibnamefont
			{Konar}}, \bibinfo {author} {\bibfnamefont {S.}~\bibnamefont {Ghosh}},\ and\
		\bibinfo {author} {\bibfnamefont {A.}~\bibnamefont {Sen(De)}},\ }\bibfield
	{title} {\bibinfo {title} {Refrigeration via purification through repeated
			measurements},\ }\href {https://doi.org/10.1103/PhysRevA.106.022616}
	{\bibfield  {journal} {\bibinfo  {journal} {Phys. Rev. A}\ }\textbf {\bibinfo
			{volume} {106}},\ \bibinfo {pages} {022616} (\bibinfo {year}
		{2022})}\BibitemShut {NoStop}%
	\bibitem [{\citenamefont {Yan}\ and\ \citenamefont {Jing}(2022)}]{YanJing2022}%
	\BibitemOpen
	\bibfield  {author} {\bibinfo {author} {\bibfnamefont {J.-s.}\ \bibnamefont
			{Yan}}\ and\ \bibinfo {author} {\bibfnamefont {J.}~\bibnamefont {Jing}},\
	}\bibfield  {title} {\bibinfo {title} {Optimizing measurement-based cooling
			by reinforcement learning},\ }\href
	{https://doi.org/10.1103/PhysRevA.106.033124} {\bibfield  {journal} {\bibinfo
			{journal} {Phys. Rev. A}\ }\textbf {\bibinfo {volume} {106}},\ \bibinfo
		{pages} {033124} (\bibinfo {year} {2022})}\BibitemShut {NoStop}%
	\bibitem [{\citenamefont {Colnaghi}\ \emph {et~al.}(2012)\citenamefont
		{Colnaghi}, \citenamefont {D'Ariano}, \citenamefont {Facchini},\ and\
		\citenamefont {Perinotti}}]{Colnaghietal2012}%
	\BibitemOpen
	\bibfield  {author} {\bibinfo {author} {\bibfnamefont {T.}~\bibnamefont
			{Colnaghi}}, \bibinfo {author} {\bibfnamefont {G.~M.}\ \bibnamefont
			{D'Ariano}}, \bibinfo {author} {\bibfnamefont {S.}~\bibnamefont {Facchini}},\
		and\ \bibinfo {author} {\bibfnamefont {P.}~\bibnamefont {Perinotti}},\
	}\bibfield  {title} {\bibinfo {title} {Quantum computation with programmable
			connections between gates},\ }\href
	{https://doi.org/https://doi.org/10.1016/j.physleta.2012.08.028} {\bibfield
		{journal} {\bibinfo  {journal} {Physics Letters A}\ }\textbf {\bibinfo
			{volume} {376}},\ \bibinfo {pages} {2940} (\bibinfo {year}
		{2012})}\BibitemShut {NoStop}%
	\bibitem [{\citenamefont {Morimae}(2014)}]{Morimae2014}%
	\BibitemOpen
	\bibfield  {author} {\bibinfo {author} {\bibfnamefont {T.}~\bibnamefont
			{Morimae}},\ }\bibfield  {title} {\bibinfo {title} {Acausal measurement-based
			quantum computing},\ }\href {https://doi.org/10.1103/PhysRevA.90.010101}
	{\bibfield  {journal} {\bibinfo  {journal} {Phys. Rev. A}\ }\textbf {\bibinfo
			{volume} {90}},\ \bibinfo {pages} {010101} (\bibinfo {year}
		{2014})}\BibitemShut {NoStop}%
	\bibitem [{\citenamefont {Ara\'ujo}\ \emph {et~al.}(2014)\citenamefont
		{Ara\'ujo}, \citenamefont {Costa},\ and\ \citenamefont
		{Brukner}}]{Araujoetal2014}%
	\BibitemOpen
	\bibfield  {author} {\bibinfo {author} {\bibfnamefont {M.}~\bibnamefont
			{Ara\'ujo}}, \bibinfo {author} {\bibfnamefont {F.}~\bibnamefont {Costa}},\
		and\ \bibinfo {author} {\bibfnamefont {{\v{C}}.}~\bibnamefont {Brukner}},\
	}\bibfield  {title} {\bibinfo {title} {Computational advantage from
			quantum-controlled ordering of gates},\ }\href
	{https://doi.org/10.1103/PhysRevLett.113.250402} {\bibfield  {journal}
		{\bibinfo  {journal} {Phys. Rev. Lett.}\ }\textbf {\bibinfo {volume} {113}},\
		\bibinfo {pages} {250402} (\bibinfo {year} {2014})}\BibitemShut {NoStop}%
	\bibitem [{\citenamefont {Taddei}\ \emph {et~al.}(2021)\citenamefont {Taddei},
		\citenamefont {Cari\~ne}, \citenamefont {Mart\'{\i}nez}, \citenamefont
		{Garc\'{\i}a}, \citenamefont {Guerrero}, \citenamefont {Abbott},
		\citenamefont {Ara\'ujo}, \citenamefont {Branciard}, \citenamefont {G\'omez},
		\citenamefont {Walborn}, \citenamefont {Aolita},\ and\ \citenamefont
		{Lima}}]{Taddeietal2021}%
	\BibitemOpen
	\bibfield  {author} {\bibinfo {author} {\bibfnamefont {M.~M.}\ \bibnamefont
			{Taddei}}, \bibinfo {author} {\bibfnamefont {J.}~\bibnamefont {Cari\~ne}},
		\bibinfo {author} {\bibfnamefont {D.}~\bibnamefont {Mart\'{\i}nez}}, \bibinfo
		{author} {\bibfnamefont {T.}~\bibnamefont {Garc\'{\i}a}}, \bibinfo {author}
		{\bibfnamefont {N.}~\bibnamefont {Guerrero}}, \bibinfo {author}
		{\bibfnamefont {A.~A.}\ \bibnamefont {Abbott}}, \bibinfo {author}
		{\bibfnamefont {M.}~\bibnamefont {Ara\'ujo}}, \bibinfo {author}
		{\bibfnamefont {C.}~\bibnamefont {Branciard}}, \bibinfo {author}
		{\bibfnamefont {E.~S.}\ \bibnamefont {G\'omez}}, \bibinfo {author}
		{\bibfnamefont {S.~P.}\ \bibnamefont {Walborn}}, \bibinfo {author}
		{\bibfnamefont {L.}~\bibnamefont {Aolita}},\ and\ \bibinfo {author}
		{\bibfnamefont {G.}~\bibnamefont {Lima}},\ }\bibfield  {title} {\bibinfo
		{title} {Computational advantage from the quantum superposition of multiple
			temporal orders of photonic gates},\ }\href
	{https://doi.org/10.1103/PRXQuantum.2.010320} {\bibfield  {journal} {\bibinfo
			{journal} {PRX Quantum}\ }\textbf {\bibinfo {volume} {2}},\ \bibinfo {pages}
		{010320} (\bibinfo {year} {2021})}\BibitemShut {NoStop}%
	\bibitem [{\citenamefont {Chiribella}(2012)}]{Chiribella2012}%
	\BibitemOpen
	\bibfield  {author} {\bibinfo {author} {\bibfnamefont {G.}~\bibnamefont
			{Chiribella}},\ }\bibfield  {title} {\bibinfo {title} {Perfect discrimination
			of no-signalling channels via quantum superposition of causal structures},\
	}\href {https://doi.org/10.1103/PhysRevA.86.040301} {\bibfield  {journal}
		{\bibinfo  {journal} {Phys. Rev. A}\ }\textbf {\bibinfo {volume} {86}},\
		\bibinfo {pages} {040301} (\bibinfo {year} {2012})}\BibitemShut {NoStop}%
	\bibitem [{\citenamefont {Feix}\ \emph {et~al.}(2015)\citenamefont {Feix},
		\citenamefont {Ara\'ujo},\ and\ \citenamefont {Brukner}}]{Feixetal2015}%
	\BibitemOpen
	\bibfield  {author} {\bibinfo {author} {\bibfnamefont {A.}~\bibnamefont
			{Feix}}, \bibinfo {author} {\bibfnamefont {M.}~\bibnamefont {Ara\'ujo}},\
		and\ \bibinfo {author} {\bibfnamefont {{\v{C}}.}~\bibnamefont {Brukner}},\
	}\bibfield  {title} {\bibinfo {title} {Quantum superposition of the order of
			parties as a communication resource},\ }\href
	{https://doi.org/10.1103/PhysRevA.92.052326} {\bibfield  {journal} {\bibinfo
			{journal} {Phys. Rev. A}\ }\textbf {\bibinfo {volume} {92}},\ \bibinfo
		{pages} {052326} (\bibinfo {year} {2015})}\BibitemShut {NoStop}%
	\bibitem [{\citenamefont {Gu\'erin}\ \emph {et~al.}(2016)\citenamefont
		{Gu\'erin}, \citenamefont {Feix}, \citenamefont {Ara\'ujo},\ and\
		\citenamefont {Brukner}}]{Guerinetal2016}%
	\BibitemOpen
	\bibfield  {author} {\bibinfo {author} {\bibfnamefont {P.~A.}\ \bibnamefont
			{Gu\'erin}}, \bibinfo {author} {\bibfnamefont {A.}~\bibnamefont {Feix}},
		\bibinfo {author} {\bibfnamefont {M.}~\bibnamefont {Ara\'ujo}},\ and\
		\bibinfo {author} {\bibfnamefont {{\v{C}}.}~\bibnamefont {Brukner}},\
	}\bibfield  {title} {\bibinfo {title} {Exponential communication complexity
			advantage from quantum superposition of the direction of communication},\
	}\href {https://doi.org/10.1103/PhysRevLett.117.100502} {\bibfield  {journal}
		{\bibinfo  {journal} {Phys. Rev. Lett.}\ }\textbf {\bibinfo {volume} {117}},\
		\bibinfo {pages} {100502} (\bibinfo {year} {2016})}\BibitemShut {NoStop}%
	\bibitem [{\citenamefont {Del~Santo}\ and\ \citenamefont
		{Daki\ifmmode~\acute{c}\else \'{c}\fi{}}(2018)}]{DelSantoDakic2018}%
	\BibitemOpen
	\bibfield  {author} {\bibinfo {author} {\bibfnamefont {F.}~\bibnamefont
			{Del~Santo}}\ and\ \bibinfo {author} {\bibfnamefont {B.}~\bibnamefont
			{Daki\ifmmode~\acute{c}\else \'{c}\fi{}}},\ }\bibfield  {title} {\bibinfo
		{title} {Two-way communication with a single quantum particle},\ }\href
	{https://doi.org/10.1103/PhysRevLett.120.060503} {\bibfield  {journal}
		{\bibinfo  {journal} {Phys. Rev. Lett.}\ }\textbf {\bibinfo {volume} {120}},\
		\bibinfo {pages} {060503} (\bibinfo {year} {2018})}\BibitemShut {NoStop}%
	\bibitem [{\citenamefont {Ebler}\ \emph {et~al.}(2018)\citenamefont {Ebler},
		\citenamefont {Salek},\ and\ \citenamefont {Chiribella}}]{Ebleretal2018}%
	\BibitemOpen
	\bibfield  {author} {\bibinfo {author} {\bibfnamefont {D.}~\bibnamefont
			{Ebler}}, \bibinfo {author} {\bibfnamefont {S.}~\bibnamefont {Salek}},\ and\
		\bibinfo {author} {\bibfnamefont {G.}~\bibnamefont {Chiribella}},\ }\bibfield
	{title} {\bibinfo {title} {Enhanced communication with the assistance of
			indefinite causal order},\ }\href
	{https://doi.org/10.1103/PhysRevLett.120.120502} {\bibfield  {journal}
		{\bibinfo  {journal} {Phys. Rev. Lett.}\ }\textbf {\bibinfo {volume} {120}},\
		\bibinfo {pages} {120502} (\bibinfo {year} {2018})}\BibitemShut {NoStop}%
	\bibitem [{\citenamefont {Procopio}\ \emph {et~al.}(2019)\citenamefont
		{Procopio}, \citenamefont {Delgado}, \citenamefont {Enríquez}, \citenamefont
		{Belabas},\ and\ \citenamefont {Levenson}}]{Procopioetal2019}%
	\BibitemOpen
	\bibfield  {author} {\bibinfo {author} {\bibfnamefont {L.~M.}\ \bibnamefont
			{Procopio}}, \bibinfo {author} {\bibfnamefont {F.}~\bibnamefont {Delgado}},
		\bibinfo {author} {\bibfnamefont {M.}~\bibnamefont {Enríquez}}, \bibinfo
		{author} {\bibfnamefont {N.}~\bibnamefont {Belabas}},\ and\ \bibinfo {author}
		{\bibfnamefont {J.~A.}\ \bibnamefont {Levenson}},\ }\bibfield  {title}
	{\bibinfo {title} {Communication enhancement through quantum coherent control
			of n channels in an indefinite causal-order scenario},\ }\href
	{https://doi.org/10.3390/e21101012} {\bibfield  {journal} {\bibinfo
			{journal} {Entropy}\ }\textbf {\bibinfo {volume} {21}},\ \bibinfo {pages}
		{1012} (\bibinfo {year} {2019})}\BibitemShut {NoStop}%
	\bibitem [{\citenamefont {Chiribella}\ \emph {et~al.}(2021)\citenamefont
		{Chiribella}, \citenamefont {Banik}, \citenamefont {Bhattacharya},
		\citenamefont {Guha}, \citenamefont {Alimuddin}, \citenamefont {Roy},
		\citenamefont {Saha}, \citenamefont {Agrawal},\ and\ \citenamefont
		{Kar}}]{Chiribellaetal2021}%
	\BibitemOpen
	\bibfield  {author} {\bibinfo {author} {\bibfnamefont {G.}~\bibnamefont
			{Chiribella}}, \bibinfo {author} {\bibfnamefont {M.}~\bibnamefont {Banik}},
		\bibinfo {author} {\bibfnamefont {S.~S.}\ \bibnamefont {Bhattacharya}},
		\bibinfo {author} {\bibfnamefont {T.}~\bibnamefont {Guha}}, \bibinfo {author}
		{\bibfnamefont {M.}~\bibnamefont {Alimuddin}}, \bibinfo {author}
		{\bibfnamefont {A.}~\bibnamefont {Roy}}, \bibinfo {author} {\bibfnamefont
			{S.}~\bibnamefont {Saha}}, \bibinfo {author} {\bibfnamefont {S.}~\bibnamefont
			{Agrawal}},\ and\ \bibinfo {author} {\bibfnamefont {G.}~\bibnamefont {Kar}},\
	}\bibfield  {title} {\bibinfo {title} {Indefinite causal order enables
			perfect quantum communication with zero capacity channels},\ }\href
	{https://doi.org/10.1088/1367-2630/abe7a0} {\bibfield  {journal} {\bibinfo
			{journal} {New Journal of Physics}\ }\textbf {\bibinfo {volume} {23}},\
		\bibinfo {pages} {033039} (\bibinfo {year} {2021})}\BibitemShut {NoStop}%
	\bibitem [{\citenamefont {{Mukhopadhyay}}\ \emph {et~al.}(2018)\citenamefont
		{{Mukhopadhyay}}, \citenamefont {{Gupta}},\ and\ \citenamefont
		{{Pati}}}]{Mukhopadhyayetal2018arxiv}%
	\BibitemOpen
	\bibfield  {author} {\bibinfo {author} {\bibfnamefont {C.}~\bibnamefont
			{{Mukhopadhyay}}}, \bibinfo {author} {\bibfnamefont {M.~K.}\ \bibnamefont
			{{Gupta}}},\ and\ \bibinfo {author} {\bibfnamefont {A.~K.}\ \bibnamefont
			{{Pati}}},\ }\bibfield  {title} {\bibinfo {title} {{Superposition of causal
				order as a metrological resource for quantum thermometry}},\ }\href@noop {}
	{\  (\bibinfo {year} {2018})},\ \Eprint {https://arxiv.org/abs/1812.07508}
	{arXiv:1812.07508 [quant-ph]} \BibitemShut {NoStop}%
	\bibitem [{\citenamefont {Frey}(2019)}]{Frey2019}%
	\BibitemOpen
	\bibfield  {author} {\bibinfo {author} {\bibfnamefont {M.}~\bibnamefont
			{Frey}},\ }\bibfield  {title} {\bibinfo {title} {Indefinite causal order aids
			quantum depolarizing channel identification},\ }\href
	{https://doi.org/10.1007/s11128-019-2186-9} {\bibfield  {journal} {\bibinfo
			{journal} {Quantum Information Processing}\ }\textbf {\bibinfo {volume}
			{18}},\ \bibinfo {pages} {96} (\bibinfo {year} {2019})}\BibitemShut {NoStop}%
	\bibitem [{\citenamefont {Zhao}\ \emph {et~al.}(2020)\citenamefont {Zhao},
		\citenamefont {Yang},\ and\ \citenamefont {Chiribella}}]{Zhaoetal2020}%
	\BibitemOpen
	\bibfield  {author} {\bibinfo {author} {\bibfnamefont {X.}~\bibnamefont
			{Zhao}}, \bibinfo {author} {\bibfnamefont {Y.}~\bibnamefont {Yang}},\ and\
		\bibinfo {author} {\bibfnamefont {G.}~\bibnamefont {Chiribella}},\ }\bibfield
	{title} {\bibinfo {title} {Quantum metrology with indefinite causal order},\
	}\href {https://doi.org/10.1103/PhysRevLett.124.190503} {\bibfield  {journal}
		{\bibinfo  {journal} {Phys. Rev. Lett.}\ }\textbf {\bibinfo {volume} {124}},\
		\bibinfo {pages} {190503} (\bibinfo {year} {2020})}\BibitemShut {NoStop}%
	\bibitem [{\citenamefont
		{Chapeau-Blondeau}(2021{\natexlab{a}})}]{ChapeauBlondeau2021ICO}%
	\BibitemOpen
	\bibfield  {author} {\bibinfo {author} {\bibfnamefont {F.}~\bibnamefont
			{Chapeau-Blondeau}},\ }\bibfield  {title} {\bibinfo {title} {Noisy quantum
			metrology with the assistance of indefinite causal order},\ }\href
	{https://doi.org/10.1103/PhysRevA.103.032615} {\bibfield  {journal} {\bibinfo
			{journal} {Phys. Rev. A}\ }\textbf {\bibinfo {volume} {103}},\ \bibinfo
		{pages} {032615} (\bibinfo {year} {2021}{\natexlab{a}})}\BibitemShut
	{NoStop}%
	\bibitem [{\citenamefont {Guha}\ \emph {et~al.}(2020)\citenamefont {Guha},
		\citenamefont {Alimuddin},\ and\ \citenamefont {Parashar}}]{Guhaetal2020}%
	\BibitemOpen
	\bibfield  {author} {\bibinfo {author} {\bibfnamefont {T.}~\bibnamefont
			{Guha}}, \bibinfo {author} {\bibfnamefont {M.}~\bibnamefont {Alimuddin}},\
		and\ \bibinfo {author} {\bibfnamefont {P.}~\bibnamefont {Parashar}},\
	}\bibfield  {title} {\bibinfo {title} {Thermodynamic advancement in the
			causally inseparable occurrence of thermal maps},\ }\href
	{https://doi.org/10.1103/PhysRevA.102.032215} {\bibfield  {journal} {\bibinfo
			{journal} {Phys. Rev. A}\ }\textbf {\bibinfo {volume} {102}},\ \bibinfo
		{pages} {032215} (\bibinfo {year} {2020})}\BibitemShut {NoStop}%
	\bibitem [{\citenamefont {Cao}\ \emph {et~al.}(2022)\citenamefont {Cao},
		\citenamefont {Wang}, \citenamefont {Jia}, \citenamefont {Zhang},
		\citenamefont {Guo}, \citenamefont {Liu}, \citenamefont {Huang},
		\citenamefont {Li},\ and\ \citenamefont {Guo}}]{Caoetal2022}%
	\BibitemOpen
	\bibfield  {author} {\bibinfo {author} {\bibfnamefont {H.}~\bibnamefont
			{Cao}}, \bibinfo {author} {\bibfnamefont {N.-N.}\ \bibnamefont {Wang}},
		\bibinfo {author} {\bibfnamefont {Z.}~\bibnamefont {Jia}}, \bibinfo {author}
		{\bibfnamefont {C.}~\bibnamefont {Zhang}}, \bibinfo {author} {\bibfnamefont
			{Y.}~\bibnamefont {Guo}}, \bibinfo {author} {\bibfnamefont {B.-H.}\
			\bibnamefont {Liu}}, \bibinfo {author} {\bibfnamefont {Y.-F.}\ \bibnamefont
			{Huang}}, \bibinfo {author} {\bibfnamefont {C.-F.}\ \bibnamefont {Li}},\ and\
		\bibinfo {author} {\bibfnamefont {G.-C.}\ \bibnamefont {Guo}},\ }\bibfield
	{title} {\bibinfo {title} {Quantum simulation of indefinite causal order
			induced quantum refrigeration},\ }\href
	{https://doi.org/10.1103/PhysRevResearch.4.L032029} {\bibfield  {journal}
		{\bibinfo  {journal} {Phys. Rev. Res.}\ }\textbf {\bibinfo {volume} {4}},\
		\bibinfo {pages} {L032029} (\bibinfo {year} {2022})}\BibitemShut {NoStop}%
	\bibitem [{\citenamefont {{Chen}}\ and\ \citenamefont
		{{Hasegawa}}(2021)}]{ChenHasegawa2021arxiv}%
	\BibitemOpen
	\bibfield  {author} {\bibinfo {author} {\bibfnamefont {Y.}~\bibnamefont
			{{Chen}}}\ and\ \bibinfo {author} {\bibfnamefont {Y.}~\bibnamefont
			{{Hasegawa}}},\ }\bibfield  {title} {\bibinfo {title} {{Indefinite Causal
				Order in Quantum Batteries}},\ }\href@noop {} {\  (\bibinfo {year} {2021})},\
	\Eprint {https://arxiv.org/abs/2105.12466} {arXiv:2105.12466 [quant-ph]}
	\BibitemShut {NoStop}%
	\bibitem [{\citenamefont {Guha}\ \emph {et~al.}(2022)\citenamefont {Guha},
		\citenamefont {Roy}, \citenamefont {Simonov},\ and\ \citenamefont
		{Zimborás}}]{Guhaetal2022arxiv}%
	\BibitemOpen
	\bibfield  {author} {\bibinfo {author} {\bibfnamefont {T.}~\bibnamefont
			{Guha}}, \bibinfo {author} {\bibfnamefont {S.}~\bibnamefont {Roy}}, \bibinfo
		{author} {\bibfnamefont {K.}~\bibnamefont {Simonov}},\ and\ \bibinfo {author}
		{\bibfnamefont {Z.}~\bibnamefont {Zimborás}},\ }\bibfield  {title} {\bibinfo
		{title} {Activation of thermal states by quantum switch-driven thermalization
			and its limits},\ }\href {https://arxiv.org/abs/2208.04034} {\  (\bibinfo
		{year} {2022})},\ \Eprint {https://arxiv.org/abs/2208.04034}
	{arXiv:2208.04034} \BibitemShut {NoStop}%
	\bibitem [{\citenamefont {Simonov}\ \emph {et~al.}(2022)\citenamefont
		{Simonov}, \citenamefont {Francica}, \citenamefont {Guarnieri},\ and\
		\citenamefont {Paternostro}}]{Simonovetal2022}%
	\BibitemOpen
	\bibfield  {author} {\bibinfo {author} {\bibfnamefont {K.}~\bibnamefont
			{Simonov}}, \bibinfo {author} {\bibfnamefont {G.}~\bibnamefont {Francica}},
		\bibinfo {author} {\bibfnamefont {G.}~\bibnamefont {Guarnieri}},\ and\
		\bibinfo {author} {\bibfnamefont {M.}~\bibnamefont {Paternostro}},\
	}\bibfield  {title} {\bibinfo {title} {Work extraction from coherently
			activated maps via quantum switch},\ }\href
	{https://doi.org/10.1103/PhysRevA.105.032217} {\bibfield  {journal} {\bibinfo
			{journal} {Phys. Rev. A}\ }\textbf {\bibinfo {volume} {105}},\ \bibinfo
		{pages} {032217} (\bibinfo {year} {2022})}\BibitemShut {NoStop}%
	\bibitem [{\citenamefont {Procopio}\ \emph {et~al.}(2015)\citenamefont
		{Procopio}, \citenamefont {Moqanaki}, \citenamefont {Araújo}, \citenamefont
		{Costa}, \citenamefont {Alonso~Calafell}, \citenamefont {Dowd}, \citenamefont
		{Hamel}, \citenamefont {Rozema}, \citenamefont {Brukner},\ and\ \citenamefont
		{Walther}}]{Procopioetal2015}%
	\BibitemOpen
	\bibfield  {author} {\bibinfo {author} {\bibfnamefont {L.~M.}\ \bibnamefont
			{Procopio}}, \bibinfo {author} {\bibfnamefont {A.}~\bibnamefont {Moqanaki}},
		\bibinfo {author} {\bibfnamefont {M.}~\bibnamefont {Araújo}}, \bibinfo
		{author} {\bibfnamefont {F.}~\bibnamefont {Costa}}, \bibinfo {author}
		{\bibfnamefont {I.}~\bibnamefont {Alonso~Calafell}}, \bibinfo {author}
		{\bibfnamefont {E.~G.}\ \bibnamefont {Dowd}}, \bibinfo {author}
		{\bibfnamefont {D.~R.}\ \bibnamefont {Hamel}}, \bibinfo {author}
		{\bibfnamefont {L.~A.}\ \bibnamefont {Rozema}}, \bibinfo {author}
		{\bibfnamefont {{\v{C}}.}~\bibnamefont {Brukner}},\ and\ \bibinfo {author}
		{\bibfnamefont {P.}~\bibnamefont {Walther}},\ }\bibfield  {title} {\bibinfo
		{title} {Experimental superposition of orders of quantum gates},\ }\href
	{https://doi.org/10.1038/ncomms8913} {\bibfield  {journal} {\bibinfo
			{journal} {Nature Communications}\ }\textbf {\bibinfo {volume} {6}},\
		\bibinfo {pages} {7913} (\bibinfo {year} {2015})}\BibitemShut {NoStop}%
	\bibitem [{\citenamefont {Rubino}\ \emph {et~al.}(2017)\citenamefont {Rubino},
		\citenamefont {Rozema}, \citenamefont {Feix}, \citenamefont {Ara{\'u}jo},
		\citenamefont {Zeuner}, \citenamefont {Procopio}, \citenamefont {Brukner},\
		and\ \citenamefont {Walther}}]{Rubinoetal2017}%
	\BibitemOpen
	\bibfield  {author} {\bibinfo {author} {\bibfnamefont {G.}~\bibnamefont
			{Rubino}}, \bibinfo {author} {\bibfnamefont {L.~A.}\ \bibnamefont {Rozema}},
		\bibinfo {author} {\bibfnamefont {A.}~\bibnamefont {Feix}}, \bibinfo {author}
		{\bibfnamefont {M.}~\bibnamefont {Ara{\'u}jo}}, \bibinfo {author}
		{\bibfnamefont {J.~M.}\ \bibnamefont {Zeuner}}, \bibinfo {author}
		{\bibfnamefont {L.~M.}\ \bibnamefont {Procopio}}, \bibinfo {author}
		{\bibfnamefont {{\v {C}}.}~\bibnamefont {Brukner}},\ and\ \bibinfo {author}
		{\bibfnamefont {P.}~\bibnamefont {Walther}},\ }\bibfield  {title} {\bibinfo
		{title} {Experimental verification of an indefinite causal order},\
	}\bibfield  {journal} {\bibinfo  {journal} {Science Advances}\ }\textbf
	{\bibinfo {volume} {3}},\ \href {https://doi.org/10.1126/sciadv.1602589}
	{10.1126/sciadv.1602589} (\bibinfo {year} {2017})\BibitemShut {NoStop}%
	\bibitem [{\citenamefont {Rubino}\ \emph {et~al.}(2022)\citenamefont {Rubino},
		\citenamefont {Rozema}, \citenamefont {Massa}, \citenamefont {Ara{\'{u}}jo},
		\citenamefont {Zych}, \citenamefont {Brukner},\ and\ \citenamefont
		{Walther}}]{Rubinoetal2022}%
	\BibitemOpen
	\bibfield  {author} {\bibinfo {author} {\bibfnamefont {G.}~\bibnamefont
			{Rubino}}, \bibinfo {author} {\bibfnamefont {L.~A.}\ \bibnamefont {Rozema}},
		\bibinfo {author} {\bibfnamefont {F.}~\bibnamefont {Massa}}, \bibinfo
		{author} {\bibfnamefont {M.}~\bibnamefont {Ara{\'{u}}jo}}, \bibinfo {author}
		{\bibfnamefont {M.}~\bibnamefont {Zych}}, \bibinfo {author} {\bibfnamefont
			{{\v{C}}.}~\bibnamefont {Brukner}},\ and\ \bibinfo {author} {\bibfnamefont
			{P.}~\bibnamefont {Walther}},\ }\bibfield  {title} {\bibinfo {title}
		{Experimental entanglement of temporal order},\ }\href
	{https://doi.org/10.22331/q-2022-01-11-621} {\bibfield  {journal} {\bibinfo
			{journal} {{Quantum}}\ }\textbf {\bibinfo {volume} {6}},\ \bibinfo {pages}
		{621} (\bibinfo {year} {2022})}\BibitemShut {NoStop}%
	\bibitem [{\citenamefont {Goswami}\ \emph {et~al.}(2018)\citenamefont
		{Goswami}, \citenamefont {Giarmatzi}, \citenamefont {Kewming}, \citenamefont
		{Costa}, \citenamefont {Branciard}, \citenamefont {Romero},\ and\
		\citenamefont {White}}]{Goswamietal2018}%
	\BibitemOpen
	\bibfield  {author} {\bibinfo {author} {\bibfnamefont {K.}~\bibnamefont
			{Goswami}}, \bibinfo {author} {\bibfnamefont {C.}~\bibnamefont {Giarmatzi}},
		\bibinfo {author} {\bibfnamefont {M.}~\bibnamefont {Kewming}}, \bibinfo
		{author} {\bibfnamefont {F.}~\bibnamefont {Costa}}, \bibinfo {author}
		{\bibfnamefont {C.}~\bibnamefont {Branciard}}, \bibinfo {author}
		{\bibfnamefont {J.}~\bibnamefont {Romero}},\ and\ \bibinfo {author}
		{\bibfnamefont {A.~G.}\ \bibnamefont {White}},\ }\bibfield  {title} {\bibinfo
		{title} {Indefinite causal order in a quantum switch},\ }\href
	{https://doi.org/10.1103/PhysRevLett.121.090503} {\bibfield  {journal}
		{\bibinfo  {journal} {Phys. Rev. Lett.}\ }\textbf {\bibinfo {volume} {121}},\
		\bibinfo {pages} {090503} (\bibinfo {year} {2018})}\BibitemShut {NoStop}%
	\bibitem [{\citenamefont {Wei}\ \emph {et~al.}(2019)\citenamefont {Wei},
		\citenamefont {Tischler}, \citenamefont {Zhao}, \citenamefont {Li},
		\citenamefont {Arrazola}, \citenamefont {Liu}, \citenamefont {Zhang},
		\citenamefont {Li}, \citenamefont {You}, \citenamefont {Wang}, \citenamefont
		{Chen}, \citenamefont {Sanders}, \citenamefont {Zhang}, \citenamefont
		{Pryde}, \citenamefont {Xu},\ and\ \citenamefont {Pan}}]{Weietal2019}%
	\BibitemOpen
	\bibfield  {author} {\bibinfo {author} {\bibfnamefont {K.}~\bibnamefont
			{Wei}}, \bibinfo {author} {\bibfnamefont {N.}~\bibnamefont {Tischler}},
		\bibinfo {author} {\bibfnamefont {S.-R.}\ \bibnamefont {Zhao}}, \bibinfo
		{author} {\bibfnamefont {Y.-H.}\ \bibnamefont {Li}}, \bibinfo {author}
		{\bibfnamefont {J.~M.}\ \bibnamefont {Arrazola}}, \bibinfo {author}
		{\bibfnamefont {Y.}~\bibnamefont {Liu}}, \bibinfo {author} {\bibfnamefont
			{W.}~\bibnamefont {Zhang}}, \bibinfo {author} {\bibfnamefont
			{H.}~\bibnamefont {Li}}, \bibinfo {author} {\bibfnamefont {L.}~\bibnamefont
			{You}}, \bibinfo {author} {\bibfnamefont {Z.}~\bibnamefont {Wang}}, \bibinfo
		{author} {\bibfnamefont {Y.-A.}\ \bibnamefont {Chen}}, \bibinfo {author}
		{\bibfnamefont {B.~C.}\ \bibnamefont {Sanders}}, \bibinfo {author}
		{\bibfnamefont {Q.}~\bibnamefont {Zhang}}, \bibinfo {author} {\bibfnamefont
			{G.~J.}\ \bibnamefont {Pryde}}, \bibinfo {author} {\bibfnamefont
			{F.}~\bibnamefont {Xu}},\ and\ \bibinfo {author} {\bibfnamefont {J.-W.}\
			\bibnamefont {Pan}},\ }\bibfield  {title} {\bibinfo {title} {Experimental
			quantum switching for exponentially superior quantum communication
			complexity},\ }\href {https://doi.org/10.1103/PhysRevLett.122.120504}
	{\bibfield  {journal} {\bibinfo  {journal} {Phys. Rev. Lett.}\ }\textbf
		{\bibinfo {volume} {122}},\ \bibinfo {pages} {120504} (\bibinfo {year}
		{2019})}\BibitemShut {NoStop}%
	\bibitem [{\citenamefont {Massa}\ \emph {et~al.}(2019)\citenamefont {Massa},
		\citenamefont {Moqanaki}, \citenamefont {Baumeler}, \citenamefont
		{Del~Santo}, \citenamefont {Kettlewell}, \citenamefont {Dakić},\ and\
		\citenamefont {Walther}}]{Massaetal2019}%
	\BibitemOpen
	\bibfield  {author} {\bibinfo {author} {\bibfnamefont {F.}~\bibnamefont
			{Massa}}, \bibinfo {author} {\bibfnamefont {A.}~\bibnamefont {Moqanaki}},
		\bibinfo {author} {\bibfnamefont {{\"A}.}~\bibnamefont {Baumeler}}, \bibinfo
		{author} {\bibfnamefont {F.}~\bibnamefont {Del~Santo}}, \bibinfo {author}
		{\bibfnamefont {J.~A.}\ \bibnamefont {Kettlewell}}, \bibinfo {author}
		{\bibfnamefont {B.}~\bibnamefont {Dakić}},\ and\ \bibinfo {author}
		{\bibfnamefont {P.}~\bibnamefont {Walther}},\ }\bibfield  {title} {\bibinfo
		{title} {Experimental two-way communication with one photon},\ }\href
	{https://doi.org/https://doi.org/10.1002/qute.201900050} {\bibfield
		{journal} {\bibinfo  {journal} {Advanced Quantum Technologies}\ }\textbf
		{\bibinfo {volume} {2}},\ \bibinfo {pages} {1900050} (\bibinfo {year}
		{2019})}\BibitemShut {NoStop}%
	\bibitem [{\citenamefont {Goswami}\ \emph {et~al.}(2020)\citenamefont
		{Goswami}, \citenamefont {Cao}, \citenamefont {Paz-Silva}, \citenamefont
		{Romero},\ and\ \citenamefont {White}}]{Goswamietal2020}%
	\BibitemOpen
	\bibfield  {author} {\bibinfo {author} {\bibfnamefont {K.}~\bibnamefont
			{Goswami}}, \bibinfo {author} {\bibfnamefont {Y.}~\bibnamefont {Cao}},
		\bibinfo {author} {\bibfnamefont {G.~A.}\ \bibnamefont {Paz-Silva}}, \bibinfo
		{author} {\bibfnamefont {J.}~\bibnamefont {Romero}},\ and\ \bibinfo {author}
		{\bibfnamefont {A.~G.}\ \bibnamefont {White}},\ }\bibfield  {title} {\bibinfo
		{title} {Increasing communication capacity via superposition of order},\
	}\href {https://doi.org/10.1103/PhysRevResearch.2.033292} {\bibfield
		{journal} {\bibinfo  {journal} {Phys. Rev. Research}\ }\textbf {\bibinfo
			{volume} {2}},\ \bibinfo {pages} {033292} (\bibinfo {year}
		{2020})}\BibitemShut {NoStop}%
	\bibitem [{\citenamefont {Guo}\ \emph {et~al.}(2020)\citenamefont {Guo},
		\citenamefont {Hu}, \citenamefont {Hou}, \citenamefont {Cao}, \citenamefont
		{Cui}, \citenamefont {Liu}, \citenamefont {Huang}, \citenamefont {Li},
		\citenamefont {Guo},\ and\ \citenamefont {Chiribella}}]{Guoetal2020}%
	\BibitemOpen
	\bibfield  {author} {\bibinfo {author} {\bibfnamefont {Y.}~\bibnamefont
			{Guo}}, \bibinfo {author} {\bibfnamefont {X.-M.}\ \bibnamefont {Hu}},
		\bibinfo {author} {\bibfnamefont {Z.-B.}\ \bibnamefont {Hou}}, \bibinfo
		{author} {\bibfnamefont {H.}~\bibnamefont {Cao}}, \bibinfo {author}
		{\bibfnamefont {J.-M.}\ \bibnamefont {Cui}}, \bibinfo {author} {\bibfnamefont
			{B.-H.}\ \bibnamefont {Liu}}, \bibinfo {author} {\bibfnamefont {Y.-F.}\
			\bibnamefont {Huang}}, \bibinfo {author} {\bibfnamefont {C.-F.}\ \bibnamefont
			{Li}}, \bibinfo {author} {\bibfnamefont {G.-C.}\ \bibnamefont {Guo}},\ and\
		\bibinfo {author} {\bibfnamefont {G.}~\bibnamefont {Chiribella}},\ }\bibfield
	{title} {\bibinfo {title} {Experimental transmission of quantum information
			using a superposition of causal orders},\ }\href
	{https://doi.org/10.1103/PhysRevLett.124.030502} {\bibfield  {journal}
		{\bibinfo  {journal} {Phys. Rev. Lett.}\ }\textbf {\bibinfo {volume} {124}},\
		\bibinfo {pages} {030502} (\bibinfo {year} {2020})}\BibitemShut {NoStop}%
	\bibitem [{\citenamefont {Rubino}\ \emph {et~al.}(2021)\citenamefont {Rubino},
		\citenamefont {Rozema}, \citenamefont {Ebler}, \citenamefont
		{Kristj\'ansson}, \citenamefont {Salek}, \citenamefont {Allard~Gu\'erin},
		\citenamefont {Abbott}, \citenamefont {Branciard}, \citenamefont {Brukner},
		\citenamefont {Chiribella},\ and\ \citenamefont {Walther}}]{Rubinoetal2021}%
	\BibitemOpen
	\bibfield  {author} {\bibinfo {author} {\bibfnamefont {G.}~\bibnamefont
			{Rubino}}, \bibinfo {author} {\bibfnamefont {L.~A.}\ \bibnamefont {Rozema}},
		\bibinfo {author} {\bibfnamefont {D.}~\bibnamefont {Ebler}}, \bibinfo
		{author} {\bibfnamefont {H.}~\bibnamefont {Kristj\'ansson}}, \bibinfo
		{author} {\bibfnamefont {S.}~\bibnamefont {Salek}}, \bibinfo {author}
		{\bibfnamefont {P.}~\bibnamefont {Allard~Gu\'erin}}, \bibinfo {author}
		{\bibfnamefont {A.~A.}\ \bibnamefont {Abbott}}, \bibinfo {author}
		{\bibfnamefont {C.}~\bibnamefont {Branciard}}, \bibinfo {author}
		{\bibfnamefont {{\v{C}}.}~\bibnamefont {Brukner}}, \bibinfo {author}
		{\bibfnamefont {G.}~\bibnamefont {Chiribella}},\ and\ \bibinfo {author}
		{\bibfnamefont {P.}~\bibnamefont {Walther}},\ }\bibfield  {title} {\bibinfo
		{title} {Experimental quantum communication enhancement by superposing
			trajectories},\ }\href {https://doi.org/10.1103/PhysRevResearch.3.013093}
	{\bibfield  {journal} {\bibinfo  {journal} {Phys. Rev. Research}\ }\textbf
		{\bibinfo {volume} {3}},\ \bibinfo {pages} {013093} (\bibinfo {year}
		{2021})}\BibitemShut {NoStop}%
	\bibitem [{\citenamefont
		{Chapeau-Blondeau}(2021{\natexlab{b}})}]{ChapeauBlondeau2021ICOinspired}%
	\BibitemOpen
	\bibfield  {author} {\bibinfo {author} {\bibfnamefont {F.~m.~c.}\
			\bibnamefont {Chapeau-Blondeau}},\ }\bibfield  {title} {\bibinfo {title}
		{Quantum parameter estimation on coherently superposed noisy channels},\
	}\href {https://doi.org/10.1103/PhysRevA.104.032214} {\bibfield  {journal}
		{\bibinfo  {journal} {Phys. Rev. A}\ }\textbf {\bibinfo {volume} {104}},\
		\bibinfo {pages} {032214} (\bibinfo {year} {2021}{\natexlab{b}})}\BibitemShut
	{NoStop}%
	\bibitem [{\citenamefont {Nie}\ \emph {et~al.}(2022{\natexlab{a}})\citenamefont
		{Nie}, \citenamefont {Feng}, \citenamefont {Longden},\ and\ \citenamefont
		{Vedral}}]{Nieetal2022arxiv}%
	\BibitemOpen
	\bibfield  {author} {\bibinfo {author} {\bibfnamefont {H.}~\bibnamefont
			{Nie}}, \bibinfo {author} {\bibfnamefont {T.}~\bibnamefont {Feng}}, \bibinfo
		{author} {\bibfnamefont {S.}~\bibnamefont {Longden}},\ and\ \bibinfo {author}
		{\bibfnamefont {V.}~\bibnamefont {Vedral}},\ }\bibfield  {title} {\bibinfo
		{title} {Quantum cooling activated by coherent-controlled thermalisation},\
	}\href@noop {} {\  (\bibinfo {year} {2022}{\natexlab{a}})},\ \Eprint
	{https://arxiv.org/abs/2201.06954} {arXiv:2201.06954 [quant-ph]} \BibitemShut
	{NoStop}%
	\bibitem [{\citenamefont {{Chiribella}}\ and\ \citenamefont
		{{Zhao}}(2022)}]{ChiribellaZhao2022arxiv}%
	\BibitemOpen
	\bibfield  {author} {\bibinfo {author} {\bibfnamefont {G.}~\bibnamefont
			{{Chiribella}}}\ and\ \bibinfo {author} {\bibfnamefont {X.}~\bibnamefont
			{{Zhao}}},\ }\bibfield  {title} {\bibinfo {title} {{Heisenberg-limited
				metrology with coherent control on the probes' configuration}},\ }\href@noop
	{} {\  (\bibinfo {year} {2022})},\ \Eprint {https://arxiv.org/abs/2206.03052}
	{arXiv:2206.03052 [quant-ph]} \BibitemShut {NoStop}%
	\bibitem [{\citenamefont {Nie}\ \emph {et~al.}(2022{\natexlab{b}})\citenamefont
		{Nie}, \citenamefont {Zhu}, \citenamefont {Huang}, \citenamefont {Tang},
		\citenamefont {Long}, \citenamefont {Lin}, \citenamefont {Tian},
		\citenamefont {Qiu}, \citenamefont {Xi}, \citenamefont {Yang}, \citenamefont
		{Li}, \citenamefont {Dong}, \citenamefont {Xin},\ and\ \citenamefont
		{Lu}}]{Nieetal2022inPress}%
	\BibitemOpen
	\bibfield  {author} {\bibinfo {author} {\bibfnamefont {X.}~\bibnamefont
			{Nie}}, \bibinfo {author} {\bibfnamefont {X.}~\bibnamefont {Zhu}}, \bibinfo
		{author} {\bibfnamefont {K.}~\bibnamefont {Huang}}, \bibinfo {author}
		{\bibfnamefont {K.}~\bibnamefont {Tang}}, \bibinfo {author} {\bibfnamefont
			{X.}~\bibnamefont {Long}}, \bibinfo {author} {\bibfnamefont {Z.}~\bibnamefont
			{Lin}}, \bibinfo {author} {\bibfnamefont {Y.}~\bibnamefont {Tian}}, \bibinfo
		{author} {\bibfnamefont {C.}~\bibnamefont {Qiu}}, \bibinfo {author}
		{\bibfnamefont {C.}~\bibnamefont {Xi}}, \bibinfo {author} {\bibfnamefont
			{X.}~\bibnamefont {Yang}}, \bibinfo {author} {\bibfnamefont {J.}~\bibnamefont
			{Li}}, \bibinfo {author} {\bibfnamefont {Y.}~\bibnamefont {Dong}}, \bibinfo
		{author} {\bibfnamefont {T.}~\bibnamefont {Xin}},\ and\ \bibinfo {author}
		{\bibfnamefont {D.}~\bibnamefont {Lu}},\ }\bibfield  {title} {\bibinfo
		{title} {Experimental realization of a quantum refrigerator driven by
			indefinite causal orders},\ }\href
	{https://journals.aps.org/prl/accepted/0207cY30M5a18b61e3049f436a7c4f132032e67c7}
	{\bibfield  {journal} {\bibinfo  {journal} {Physical Review Letters}\ }
		(\bibinfo {year} {2022}{\natexlab{b}})}\BibitemShut {NoStop}%
	\bibitem [{\citenamefont {Barrett}\ \emph {et~al.}(2003)\citenamefont
		{Barrett}, \citenamefont {DeMarco}, \citenamefont {Schaetz}, \citenamefont
		{Meyer}, \citenamefont {Leibfried}, \citenamefont {Britton}, \citenamefont
		{Chiaverini}, \citenamefont {Itano}, \citenamefont
		{Jelenkovi\ifmmode~\acute{c}\else \'{c}\fi{}}, \citenamefont {Jost},
		\citenamefont {Langer}, \citenamefont {Rosenband},\ and\ \citenamefont
		{Wineland}}]{Barrettetal2003}%
	\BibitemOpen
	\bibfield  {author} {\bibinfo {author} {\bibfnamefont {M.~D.}\ \bibnamefont
			{Barrett}}, \bibinfo {author} {\bibfnamefont {B.}~\bibnamefont {DeMarco}},
		\bibinfo {author} {\bibfnamefont {T.}~\bibnamefont {Schaetz}}, \bibinfo
		{author} {\bibfnamefont {V.}~\bibnamefont {Meyer}}, \bibinfo {author}
		{\bibfnamefont {D.}~\bibnamefont {Leibfried}}, \bibinfo {author}
		{\bibfnamefont {J.}~\bibnamefont {Britton}}, \bibinfo {author} {\bibfnamefont
			{J.}~\bibnamefont {Chiaverini}}, \bibinfo {author} {\bibfnamefont {W.~M.}\
			\bibnamefont {Itano}}, \bibinfo {author} {\bibfnamefont {B.}~\bibnamefont
			{Jelenkovi\ifmmode~\acute{c}\else \'{c}\fi{}}}, \bibinfo {author}
		{\bibfnamefont {J.~D.}\ \bibnamefont {Jost}}, \bibinfo {author}
		{\bibfnamefont {C.}~\bibnamefont {Langer}}, \bibinfo {author} {\bibfnamefont
			{T.}~\bibnamefont {Rosenband}},\ and\ \bibinfo {author} {\bibfnamefont
			{D.~J.}\ \bibnamefont {Wineland}},\ }\bibfield  {title} {\bibinfo {title}
		{Sympathetic cooling of ${}^{9}{\mathrm{be}}^{+}$ and
			${}^{24}{\mathrm{mg}}^{+}$ for quantum logic},\ }\href
	{https://doi.org/10.1103/PhysRevA.68.042302} {\bibfield  {journal} {\bibinfo
			{journal} {Phys. Rev. A}\ }\textbf {\bibinfo {volume} {68}},\ \bibinfo
		{pages} {042302} (\bibinfo {year} {2003})}\BibitemShut {NoStop}%
	\bibitem [{\citenamefont {Wu}\ \emph {et~al.}(2007)\citenamefont {Wu},
		\citenamefont {Pechen}, \citenamefont {Brif},\ and\ \citenamefont
		{Rabitz}}]{Wuetal2007}%
	\BibitemOpen
	\bibfield  {author} {\bibinfo {author} {\bibfnamefont {R.}~\bibnamefont
			{Wu}}, \bibinfo {author} {\bibfnamefont {A.}~\bibnamefont {Pechen}}, \bibinfo
		{author} {\bibfnamefont {C.}~\bibnamefont {Brif}},\ and\ \bibinfo {author}
		{\bibfnamefont {H.}~\bibnamefont {Rabitz}},\ }\bibfield  {title} {\bibinfo
		{title} {Controllability of open quantum systems with {K}raus-map dynamics},\
	}\href {https://doi.org/10.1088/1751-8113/40/21/015} {\bibfield  {journal}
		{\bibinfo  {journal} {Journal of Physics A: Mathematical and Theoretical}\
		}\textbf {\bibinfo {volume} {40}},\ \bibinfo {pages} {5681} (\bibinfo {year}
		{2007})}\BibitemShut {NoStop}%
	\bibitem [{\citenamefont {Saeedi}\ and\ \citenamefont
		{Pedram}(2013)}]{SaeediPedram2013}%
	\BibitemOpen
	\bibfield  {author} {\bibinfo {author} {\bibfnamefont {M.}~\bibnamefont
			{Saeedi}}\ and\ \bibinfo {author} {\bibfnamefont {M.}~\bibnamefont
			{Pedram}},\ }\bibfield  {title} {\bibinfo {title} {Linear-depth quantum
			circuits for $n$-qubit toffoli gates with no ancilla},\ }\href
	{https://doi.org/10.1103/PhysRevA.87.062318} {\bibfield  {journal} {\bibinfo
			{journal} {Phys. Rev. A}\ }\textbf {\bibinfo {volume} {87}},\ \bibinfo
		{pages} {062318} (\bibinfo {year} {2013})}\BibitemShut {NoStop}%
	\bibitem [{\citenamefont {Baker}\ \emph {et~al.}(2021)\citenamefont {Baker},
		\citenamefont {Huber}, \citenamefont {Glaser}, \citenamefont {Roy},
		\citenamefont {Tsitsilin}, \citenamefont {Filipp},\ and\ \citenamefont
		{Hartmann}}]{Bakeretal2021arxiv}%
	\BibitemOpen
	\bibfield  {author} {\bibinfo {author} {\bibfnamefont {A.~J.}\ \bibnamefont
			{Baker}}, \bibinfo {author} {\bibfnamefont {G.~B.~P.}\ \bibnamefont {Huber}},
		\bibinfo {author} {\bibfnamefont {N.~J.}\ \bibnamefont {Glaser}}, \bibinfo
		{author} {\bibfnamefont {F.}~\bibnamefont {Roy}}, \bibinfo {author}
		{\bibfnamefont {I.}~\bibnamefont {Tsitsilin}}, \bibinfo {author}
		{\bibfnamefont {S.}~\bibnamefont {Filipp}},\ and\ \bibinfo {author}
		{\bibfnamefont {M.~J.}\ \bibnamefont {Hartmann}},\ }\bibfield  {title}
	{\bibinfo {title} {Single shot i-toffoli gate in dispersively coupled
			superconducting qubits},\ }\href@noop {} {\  (\bibinfo {year} {2021})},\
	\Eprint {https://arxiv.org/abs/2111.05938} {arXiv:2111.05938 [quant-ph]}
	\BibitemShut {NoStop}%
	\bibitem [{\citenamefont {Hardy}(2007)}]{Hardy2007}%
	\BibitemOpen
	\bibfield  {author} {\bibinfo {author} {\bibfnamefont {L.}~\bibnamefont
			{Hardy}},\ }\bibfield  {title} {\bibinfo {title} {Towards quantum gravity: a
			framework for probabilistic theories with non-fixed causal structure},\
	}\href {https://doi.org/10.1088/1751-8113/40/12/s12} {\bibfield  {journal}
		{\bibinfo  {journal} {Journal of Physics A: Mathematical and Theoretical}\
		}\textbf {\bibinfo {volume} {40}},\ \bibinfo {pages} {3081} (\bibinfo {year}
		{2007})}\BibitemShut {NoStop}%
	\bibitem [{\citenamefont {Oreshkov}\ \emph {et~al.}(2012)\citenamefont
		{Oreshkov}, \citenamefont {Costa},\ and\ \citenamefont
		{Brukner}}]{Oreshkovetal2012}%
	\BibitemOpen
	\bibfield  {author} {\bibinfo {author} {\bibfnamefont {O.}~\bibnamefont
			{Oreshkov}}, \bibinfo {author} {\bibfnamefont {F.}~\bibnamefont {Costa}},\
		and\ \bibinfo {author} {\bibfnamefont {{\v{C}}.}~\bibnamefont {Brukner}},\
	}\bibfield  {title} {\bibinfo {title} {Quantum correlations with no causal
			order},\ }\href {https://doi.org/10.1038/ncomms2076} {\bibfield  {journal}
		{\bibinfo  {journal} {Nature Communications}\ }\textbf {\bibinfo {volume}
			{3}},\ \bibinfo {pages} {1092} (\bibinfo {year} {2012})}\BibitemShut
	{NoStop}%
	\bibitem [{\citenamefont {Ara{\'{u}}jo}\ \emph {et~al.}(2015)\citenamefont
		{Ara{\'{u}}jo}, \citenamefont {Branciard}, \citenamefont {Costa},
		\citenamefont {Feix}, \citenamefont {Giarmatzi},\ and\ \citenamefont
		{Brukner}}]{Araujoetal2015}%
	\BibitemOpen
	\bibfield  {author} {\bibinfo {author} {\bibfnamefont {M.}~\bibnamefont
			{Ara{\'{u}}jo}}, \bibinfo {author} {\bibfnamefont {C.}~\bibnamefont
			{Branciard}}, \bibinfo {author} {\bibfnamefont {F.}~\bibnamefont {Costa}},
		\bibinfo {author} {\bibfnamefont {A.}~\bibnamefont {Feix}}, \bibinfo {author}
		{\bibfnamefont {C.}~\bibnamefont {Giarmatzi}},\ and\ \bibinfo {author}
		{\bibfnamefont {{\v{C}}.}~\bibnamefont {Brukner}},\ }\bibfield  {title}
	{\bibinfo {title} {Witnessing causal nonseparability},\ }\href
	{https://doi.org/10.1088/1367-2630/17/10/102001} {\bibfield  {journal}
		{\bibinfo  {journal} {New Journal of Physics}\ }\textbf {\bibinfo {volume}
			{17}},\ \bibinfo {pages} {102001} (\bibinfo {year} {2015})}\BibitemShut
	{NoStop}%
	\bibitem [{\citenamefont {Ämin Baumeler}\ and\ \citenamefont
		{Wolf}(2016)}]{BaumelerWolf2016}%
	\BibitemOpen
	\bibfield  {author} {\bibinfo {author} {\bibnamefont {Ämin Baumeler}}\ and\
		\bibinfo {author} {\bibfnamefont {S.}~\bibnamefont {Wolf}},\ }\bibfield
	{title} {\bibinfo {title} {The space of logically consistent classical
			processes without causal order},\ }\href
	{https://doi.org/10.1088/1367-2630/18/1/013036} {\bibfield  {journal}
		{\bibinfo  {journal} {New Journal of Physics}\ }\textbf {\bibinfo {volume}
			{18}},\ \bibinfo {pages} {013036} (\bibinfo {year} {2016})}\BibitemShut
	{NoStop}%
	\bibitem [{\citenamefont {Oreshkov}\ and\ \citenamefont
		{Giarmatzi}(2016)}]{OreshkovGiarmatzi2016}%
	\BibitemOpen
	\bibfield  {author} {\bibinfo {author} {\bibfnamefont {O.}~\bibnamefont
			{Oreshkov}}\ and\ \bibinfo {author} {\bibfnamefont {C.}~\bibnamefont
			{Giarmatzi}},\ }\bibfield  {title} {\bibinfo {title} {Causal and causally
			separable processes},\ }\href {https://doi.org/10.1088/1367-2630/18/9/093020}
	{\bibfield  {journal} {\bibinfo  {journal} {New Journal of Physics}\ }\textbf
		{\bibinfo {volume} {18}},\ \bibinfo {pages} {093020} (\bibinfo {year}
		{2016})}\BibitemShut {NoStop}%
	\bibitem [{\citenamefont {Zych}\ \emph {et~al.}(2019)\citenamefont {Zych},
		\citenamefont {Costa}, \citenamefont {Pikovski},\ and\ \citenamefont
		{Brukner}}]{Zychetal2019}%
	\BibitemOpen
	\bibfield  {author} {\bibinfo {author} {\bibfnamefont {M.}~\bibnamefont
			{Zych}}, \bibinfo {author} {\bibfnamefont {F.}~\bibnamefont {Costa}},
		\bibinfo {author} {\bibfnamefont {I.}~\bibnamefont {Pikovski}},\ and\
		\bibinfo {author} {\bibfnamefont {{\v{C}}.}~\bibnamefont {Brukner}},\
	}\bibfield  {title} {\bibinfo {title} {Bell’s theorem for temporal order},\
	}\href {https://doi.org/10.1038/s41467-019-11579-x} {\bibfield  {journal}
		{\bibinfo  {journal} {Nature Communications}\ }\textbf {\bibinfo {volume}
			{10}},\ \bibinfo {pages} {3772} (\bibinfo {year} {2019})}\BibitemShut
	{NoStop}%
	\bibitem [{\citenamefont {Oreshkov}(2019)}]{Oreshkov2019}%
	\BibitemOpen
	\bibfield  {author} {\bibinfo {author} {\bibfnamefont {O.}~\bibnamefont
			{Oreshkov}},\ }\bibfield  {title} {\bibinfo {title} {Time-delocalized quantum
			subsystems and operations: on the existence of processes with indefinite
			causal structure in quantum mechanics},\ }\href
	{https://doi.org/10.22331/q-2019-12-02-206} {\bibfield  {journal} {\bibinfo
			{journal} {{Quantum}}\ }\textbf {\bibinfo {volume} {3}},\ \bibinfo {pages}
		{206} (\bibinfo {year} {2019})}\BibitemShut {NoStop}%
	\bibitem [{\citenamefont {Dimić}\ \emph {et~al.}(2020)\citenamefont {Dimić},
		\citenamefont {Milivojević}, \citenamefont {Go{\v{C}}anin}, \citenamefont
		{Móller},\ and\ \citenamefont {Brukner}}]{Dimicetal2020}%
	\BibitemOpen
	\bibfield  {author} {\bibinfo {author} {\bibfnamefont {A.}~\bibnamefont
			{Dimić}}, \bibinfo {author} {\bibfnamefont {M.}~\bibnamefont
			{Milivojević}}, \bibinfo {author} {\bibfnamefont {D.}~\bibnamefont
			{Go{\v{C}}anin}}, \bibinfo {author} {\bibfnamefont {N.~S.}\ \bibnamefont
			{Móller}},\ and\ \bibinfo {author} {\bibfnamefont {{\v{C}}.}~\bibnamefont
			{Brukner}},\ }\bibfield  {title} {\bibinfo {title} {Simulating indefinite
			causal order with rindler observers},\ }\href
	{https://doi.org/10.3389/fphy.2020.525333} {\bibfield  {journal} {\bibinfo
			{journal} {Frontiers in Physics}\ }\textbf {\bibinfo {volume} {8}},\ \bibinfo
		{pages} {470} (\bibinfo {year} {2020})}\BibitemShut {NoStop}%
	\bibitem [{\citenamefont {Milz}\ \emph {et~al.}(2021)\citenamefont {Milz},
		\citenamefont {Jurkschat}, \citenamefont {Pollock},\ and\ \citenamefont
		{Modi}}]{Milzetal2021}%
	\BibitemOpen
	\bibfield  {author} {\bibinfo {author} {\bibfnamefont {S.}~\bibnamefont
			{Milz}}, \bibinfo {author} {\bibfnamefont {D.}~\bibnamefont {Jurkschat}},
		\bibinfo {author} {\bibfnamefont {F.~A.}\ \bibnamefont {Pollock}},\ and\
		\bibinfo {author} {\bibfnamefont {K.}~\bibnamefont {Modi}},\ }\bibfield
	{title} {\bibinfo {title} {Delayed-choice causal order and nonclassical
			correlations},\ }\href {https://doi.org/10.1103/PhysRevResearch.3.023028}
	{\bibfield  {journal} {\bibinfo  {journal} {Phys. Rev. Research}\ }\textbf
		{\bibinfo {volume} {3}},\ \bibinfo {pages} {023028} (\bibinfo {year}
		{2021})}\BibitemShut {NoStop}%
	\bibitem [{\citenamefont {Chiribella}\ \emph {et~al.}(2013)\citenamefont
		{Chiribella}, \citenamefont {D'Ariano}, \citenamefont {Perinotti},\ and\
		\citenamefont {Valiron}}]{Chiribellaetal2013}%
	\BibitemOpen
	\bibfield  {author} {\bibinfo {author} {\bibfnamefont {G.}~\bibnamefont
			{Chiribella}}, \bibinfo {author} {\bibfnamefont {G.~M.}\ \bibnamefont
			{D'Ariano}}, \bibinfo {author} {\bibfnamefont {P.}~\bibnamefont
			{Perinotti}},\ and\ \bibinfo {author} {\bibfnamefont {B.}~\bibnamefont
			{Valiron}},\ }\bibfield  {title} {\bibinfo {title} {Quantum computations
			without definite causal structure},\ }\href
	{https://doi.org/10.1103/PhysRevA.88.022318} {\bibfield  {journal} {\bibinfo
			{journal} {Phys. Rev. A}\ }\textbf {\bibinfo {volume} {88}},\ \bibinfo
		{pages} {022318} (\bibinfo {year} {2013})}\BibitemShut {NoStop}%
	\bibitem [{\citenamefont {Friis}\ \emph {et~al.}(2014)\citenamefont {Friis},
		\citenamefont {Dunjko}, \citenamefont {D\"ur},\ and\ \citenamefont
		{Briegel}}]{Friisetal2014}%
	\BibitemOpen
	\bibfield  {author} {\bibinfo {author} {\bibfnamefont {N.}~\bibnamefont
			{Friis}}, \bibinfo {author} {\bibfnamefont {V.}~\bibnamefont {Dunjko}},
		\bibinfo {author} {\bibfnamefont {W.}~\bibnamefont {D\"ur}},\ and\ \bibinfo
		{author} {\bibfnamefont {H.~J.}\ \bibnamefont {Briegel}},\ }\bibfield
	{title} {\bibinfo {title} {Implementing quantum control for unknown
			subroutines},\ }\href {https://doi.org/10.1103/PhysRevA.89.030303} {\bibfield
		{journal} {\bibinfo  {journal} {Phys. Rev. A}\ }\textbf {\bibinfo {volume}
			{89}},\ \bibinfo {pages} {030303} (\bibinfo {year} {2014})}\BibitemShut
	{NoStop}%
	\bibitem [{\citenamefont {Friis}\ \emph {et~al.}(2015)\citenamefont {Friis},
		\citenamefont {Melnikov}, \citenamefont {Kirchmair},\ and\ \citenamefont
		{Briegel}}]{Friisetal2015}%
	\BibitemOpen
	\bibfield  {author} {\bibinfo {author} {\bibfnamefont {N.}~\bibnamefont
			{Friis}}, \bibinfo {author} {\bibfnamefont {A.~A.}\ \bibnamefont {Melnikov}},
		\bibinfo {author} {\bibfnamefont {G.}~\bibnamefont {Kirchmair}},\ and\
		\bibinfo {author} {\bibfnamefont {H.~J.}\ \bibnamefont {Briegel}},\
	}\bibfield  {title} {\bibinfo {title} {Coherent controlization using
			superconducting qubits},\ }\href {https://doi.org/10.1038/srep18036}
	{\bibfield  {journal} {\bibinfo  {journal} {Scientific Reports}\ }\textbf
		{\bibinfo {volume} {5}},\ \bibinfo {pages} {18036} (\bibinfo {year}
		{2015})}\BibitemShut {NoStop}%
	\bibitem [{\citenamefont {Rambo}\ \emph {et~al.}(2016)\citenamefont {Rambo},
		\citenamefont {Altepeter}, \citenamefont {Kumar},\ and\ \citenamefont
		{D'Ariano}}]{Ramboetal2016}%
	\BibitemOpen
	\bibfield  {author} {\bibinfo {author} {\bibfnamefont {T.~M.}\ \bibnamefont
			{Rambo}}, \bibinfo {author} {\bibfnamefont {J.~B.}\ \bibnamefont
			{Altepeter}}, \bibinfo {author} {\bibfnamefont {P.}~\bibnamefont {Kumar}},\
		and\ \bibinfo {author} {\bibfnamefont {G.~M.}\ \bibnamefont {D'Ariano}},\
	}\bibfield  {title} {\bibinfo {title} {Functional quantum computing: An
			optical approach},\ }\href {https://doi.org/10.1103/PhysRevA.93.052321}
	{\bibfield  {journal} {\bibinfo  {journal} {Phys. Rev. A}\ }\textbf {\bibinfo
			{volume} {93}},\ \bibinfo {pages} {052321} (\bibinfo {year}
		{2016})}\BibitemShut {NoStop}%
	\bibitem [{\citenamefont {Giacomini}\ \emph {et~al.}(2016)\citenamefont
		{Giacomini}, \citenamefont {Castro-Ruiz},\ and\ \citenamefont
		{Brukner}}]{Giacominietal2016}%
	\BibitemOpen
	\bibfield  {author} {\bibinfo {author} {\bibfnamefont {F.}~\bibnamefont
			{Giacomini}}, \bibinfo {author} {\bibfnamefont {E.}~\bibnamefont
			{Castro-Ruiz}},\ and\ \bibinfo {author} {\bibfnamefont {C.}~\bibnamefont
			{Brukner}},\ }\bibfield  {title} {\bibinfo {title} {Indefinite causal
			structures for continuous-variable systems},\ }\href
	{https://doi.org/10.1088/1367-2630/18/11/113026} {\bibfield  {journal}
		{\bibinfo  {journal} {New Journal of Physics}\ }\textbf {\bibinfo {volume}
			{18}},\ \bibinfo {pages} {113026} (\bibinfo {year} {2016})}\BibitemShut
	{NoStop}%
	\bibitem [{\citenamefont {Goswami}\ and\ \citenamefont
		{Romero}(2020)}]{GoswamiRomero2020}%
	\BibitemOpen
	\bibfield  {author} {\bibinfo {author} {\bibfnamefont {K.}~\bibnamefont
			{Goswami}}\ and\ \bibinfo {author} {\bibfnamefont {J.}~\bibnamefont
			{Romero}},\ }\bibfield  {title} {\bibinfo {title} {Experiments on quantum
			causality},\ }\href {https://doi.org/10.1116/5.0010747} {\bibfield  {journal}
		{\bibinfo  {journal} {AVS Quantum Science}\ }\textbf {\bibinfo {volume}
			{2}},\ \bibinfo {pages} {037101} (\bibinfo {year} {2020})}\BibitemShut
	{NoStop}%
	\bibitem [{\citenamefont {Pyshkin}\ \emph {et~al.}(2016)\citenamefont
		{Pyshkin}, \citenamefont {Luo}, \citenamefont {You},\ and\ \citenamefont
		{Wu}}]{Pyshkinetal2016}%
	\BibitemOpen
	\bibfield  {author} {\bibinfo {author} {\bibfnamefont {P.~V.}\ \bibnamefont
			{Pyshkin}}, \bibinfo {author} {\bibfnamefont {D.-W.}\ \bibnamefont {Luo}},
		\bibinfo {author} {\bibfnamefont {J.~Q.}\ \bibnamefont {You}},\ and\ \bibinfo
		{author} {\bibfnamefont {L.-A.}\ \bibnamefont {Wu}},\ }\bibfield  {title}
	{\bibinfo {title} {Ground-state cooling of quantum systems via a one-shot
			measurement},\ }\href {https://doi.org/10.1103/PhysRevA.93.032120} {\bibfield
		{journal} {\bibinfo  {journal} {Phys. Rev. A}\ }\textbf {\bibinfo {volume}
			{93}},\ \bibinfo {pages} {032120} (\bibinfo {year} {2016})}\BibitemShut
	{NoStop}%
	\bibitem [{\citenamefont {Wu}\ \emph {et~al.}(2013)\citenamefont {Wu},
		\citenamefont {Segal},\ and\ \citenamefont {Brumer}}]{Wuetal2013}%
	\BibitemOpen
	\bibfield  {author} {\bibinfo {author} {\bibfnamefont {L.-A.}\ \bibnamefont
			{Wu}}, \bibinfo {author} {\bibfnamefont {D.}~\bibnamefont {Segal}},\ and\
		\bibinfo {author} {\bibfnamefont {P.}~\bibnamefont {Brumer}},\ }\bibfield
	{title} {\bibinfo {title} {No-go theorem for ground state cooling given
			initial system-thermal bath factorization},\ }\href
	{https://doi.org/10.1038/srep01824} {\bibfield  {journal} {\bibinfo
			{journal} {Scientific Reports}\ }\textbf {\bibinfo {volume} {3}},\ \bibinfo
		{pages} {1824} (\bibinfo {year} {2013})}\BibitemShut {NoStop}%
	\bibitem [{\citenamefont {Ticozzi}\ and\ \citenamefont
		{Viola}(2014)}]{TicozziViola2014}%
	\BibitemOpen
	\bibfield  {author} {\bibinfo {author} {\bibfnamefont {F.}~\bibnamefont
			{Ticozzi}}\ and\ \bibinfo {author} {\bibfnamefont {L.}~\bibnamefont
			{Viola}},\ }\bibfield  {title} {\bibinfo {title} {Quantum resources for
			purification and cooling: fundamental limits and opportunities},\ }\href
	{https://doi.org/10.1038/srep05192} {\bibfield  {journal} {\bibinfo
			{journal} {Scientific Reports}\ }\textbf {\bibinfo {volume} {4}},\ \bibinfo
		{pages} {5192} (\bibinfo {year} {2014})}\BibitemShut {NoStop}%
	\bibitem [{\citenamefont {Silva}\ \emph {et~al.}(2016)\citenamefont {Silva},
		\citenamefont {Manzano}, \citenamefont {Skrzypczyk},\ and\ \citenamefont
		{Brunner}}]{Silvaetal2016}%
	\BibitemOpen
	\bibfield  {author} {\bibinfo {author} {\bibfnamefont {R.}~\bibnamefont
			{Silva}}, \bibinfo {author} {\bibfnamefont {G.}~\bibnamefont {Manzano}},
		\bibinfo {author} {\bibfnamefont {P.}~\bibnamefont {Skrzypczyk}},\ and\
		\bibinfo {author} {\bibfnamefont {N.}~\bibnamefont {Brunner}},\ }\bibfield
	{title} {\bibinfo {title} {Performance of autonomous quantum thermal
			machines: Hilbert space dimension as a thermodynamical resource},\ }\href
	{https://doi.org/10.1103/PhysRevE.94.032120} {\bibfield  {journal} {\bibinfo
			{journal} {Phys. Rev. E}\ }\textbf {\bibinfo {volume} {94}},\ \bibinfo
		{pages} {032120} (\bibinfo {year} {2016})}\BibitemShut {NoStop}%
	\bibitem [{Note2()}]{Note2}%
	\BibitemOpen
	\bibinfo {note} {{This latter process is only useful if the control qubit can
			be measured nondestructively such that a single pure qubit can be recycled
			for all $n$ processes.}}\BibitemShut {Stop}%
	\bibitem [{\citenamefont {Felce}\ and\ \citenamefont
		{Vedral}(2020)}]{FelceVedral2020}%
	\BibitemOpen
	\bibfield  {author} {\bibinfo {author} {\bibfnamefont {D.}~\bibnamefont
			{Felce}}\ and\ \bibinfo {author} {\bibfnamefont {V.}~\bibnamefont {Vedral}},\
	}\bibfield  {title} {\bibinfo {title} {Quantum refrigeration with indefinite
			causal order},\ }\href {https://doi.org/10.1103/PhysRevLett.125.070603}
	{\bibfield  {journal} {\bibinfo  {journal} {Phys. Rev. Lett.}\ }\textbf
		{\bibinfo {volume} {125}},\ \bibinfo {pages} {070603} (\bibinfo {year}
		{2020})}\BibitemShut {NoStop}%
	\bibitem [{\citenamefont {{Felce}}\ \emph {et~al.}(2021)\citenamefont
		{{Felce}}, \citenamefont {{Vedral}},\ and\ \citenamefont
		{{Tennie}}}]{Felceetal2021arxiv}%
	\BibitemOpen
	\bibfield  {author} {\bibinfo {author} {\bibfnamefont {D.}~\bibnamefont
			{{Felce}}}, \bibinfo {author} {\bibfnamefont {V.}~\bibnamefont {{Vedral}}},\
		and\ \bibinfo {author} {\bibfnamefont {F.}~\bibnamefont {{Tennie}}},\
	}\bibfield  {title} {\bibinfo {title} {{Refrigeration with Indefinite Causal
				Orderson a Cloud Quantum Computer}},\ }\href@noop {} {\  (\bibinfo {year}
		{2021})},\ \Eprint {https://arxiv.org/abs/2107.12413} {arXiv:2107.12413
		[quant-ph]} \BibitemShut {NoStop}%
\end{thebibliography}
\end{document}